\documentclass[a4paper,fleqn,usenatbib]{mnras}
\usepackage{newtxtext,newtxmath}
\usepackage[T1]{fontenc}
\usepackage{threeparttable}
\usepackage{ae,aecompl,multirow}
\DeclareRobustCommand{\VAN}[3]{#2}
\let\VANthebibliography\thebibliography
\def\thebibliography{\DeclareRobustCommand{\VAN}[3]{##3}\VANthebibliography}

\usepackage{amssymb}	
\usepackage{subfig}

\usepackage{graphicx}	
\usepackage{amsmath}	
\usepackage{url}

\title[Reverberation mapping of PKS 0736+017]{Spectroscopic reverberation mapping of Quasar PKS 0736+017: Broad-Line Region and Black-hole Mass}

\author[Pandey et al.]{
Shivangi Pandey$^{1}$\thanks{E-mail: Shivangipandey@aries.res.in},
Suvendu Rakshit$^{1}$\thanks{E-mail: suvenduat@gmail.com},
Jong-Hak Woo$^{2}$, and
C. S. Stalin$^{3}$ 
\\
$^{1}$Aryabhatta Research Institute of Observational Sciences, Nainital\textendash263001, Uttarakhand, India \\
$^{2}$ Astronomy Program, Department of Physics and Astronomy, Seoul National University, Seoul, 08826, Republic of Korea \\
$^{3}$ Indian Institute of Astrophysics, Block II, Koramangala, Bangalore-560034, India
}

\date{Accepted XXX. Received YYY; in original form ZZZ}
\pubyear{2022}

\begin{document}
\label{firstpage}
\pagerange{\pageref{firstpage}--\pageref{lastpage}}
\maketitle

\begin{abstract}
To understand the mass distribution and co-evolution of supermassive black holes with their host galaxy, it is crucial to measure the black hole mass of AGN. Reverberation mapping is a unique tool to estimate the black hole masses in AGN. We performed spectroscopic reverberation study using long-term monitoring data with more than 100 spectra of a radio-loud quasar PKS 0736+017 to estimate the size of the broad line region (BLR) and black hole mass. The optical spectrum shows strong H$\mathrm{\beta}$ and H$\mathrm{\gamma}$ emission lines. We generated the light curves of 5100{\AA} continuum flux ($f_{5100}$), H$\beta$, and H$\gamma$. All the light curves are found to be strongly variable with fractional variability of 69$\%$, 21$\%$, 30$\%$ for V-band, H$\beta$, and H$\gamma$ light curves, respectively. Along with the thermal contribution, non-thermal emission contributes to the estimated continuum luminosity at 5100\AA. Using different methods, e.g., CCF, {\small JAVELIN}, von-neumann, we estimated the size of the BLR, which is found to be 66.4$^{+6.0}_{-4.2}$ light days in the rest frame. The BLR size combined with the line width of H$\beta$ provides a black hole mass of 7.32$^{+0.89}_{-0.91} \times 10^{7}M_{\odot}$. The source closely follows the BLR size-luminosity relation of AGN.  
\end{abstract}




\section{Introduction}
\label{Introduction}
Active galactic nuclei (AGNs) are powered by the accretion of matter onto a central supermassive black hole (SMBH) of { mass M > 10$^{6} M_{\odot}$} \citep{2002ApJ...579..530W}. These are one of the most bright and persistent objects in the Universe ($\gtrapprox 10^{44}$ ergs s$^{-1} $), that outshine their entire host galaxy by emitting an abundant amount of radiation in a broad range of the electromagnetic region. It shows flux variation over a time scale of months to years \citep[e.g.,][]{doi:10.1146/annurev.aa.33.090195.001115,doi:10.1146/annurev.astro.35.1.445}. The emission from AGN peaks at the UV/optical region of the spectral energy distribution (SED). Due to the strong gravity of a black hole,  matter attracted toward it, spirals in and forms a disk-like structure known as an accretion disk. Surrounding this is the broad-line region (BLR), which is made of gas clouds orbiting around the SMBH with velocities of a few thousand km/s. The gas is ionized by the continuum radiation and emits broad emission lines due to Doppler broadening \citep{Urry1995}. 

The black hole mass is strongly correlated with the host galaxy properties suggesting a co-evolution of black hole and host galaxy \citep{2013ARA&A..51..511K} and subsequently studying these AGNs provides leverage to probe the growth and evolution of black holes and their host galaxy across the Universe. The mass of the black hole in AGN is challenging to measure because the bright central core overpowers the host galaxy, and the spatial resolution needed to resolve the central structure is beyond the capability of existing telescopes \citep[however, see][]{2018Natur.563..657G,GRAVITYCollaboration2020}. The black hole mass measurement of a radio-loud object is more difficult due to the non-thermal contribution from jet, which affects the optical continuum variation of AGN. 

Reverberation Mapping \citep[RM;][]{Blandford1982,Peterson_1993} uses the flux variability to estimate black hole masses and constrain the geometry and kinematics of the central engine. RM is based on the variation of line fluxes due to the variation of optical/UV photons from the accretion disk. RM is a well-known technique that has been applied so far in more than 100 objects to provide the BLR size and black hole mass \citep[e.g.,][]{Bahcall1972,Blandford1982,Peterson_1993, Peterson_1998, 1999ApJ...526..579W, Kaspi2000,Kaspi2007, Peterson_2002, Peterson_2004, Peterson2014, Greene2005, Dietrich2012, Zu2013, Bentz2009, Bentz2010, Barth2015, Du2014,Du2015, Woo2015, 2016ApJ...820...27D, Pei2017, Fausnaugh2017, 2016ApJ...825..126D, 2018ApJ...856....6D, 2016ApJ...818...30S, 2017ApJ...851...21G, 2017ApJ...847..125P, 2019ApJ...886...93R, Zhang2019, 2020ApJ...892...93C, Bonta2020, 2020A&A...642A..59R, Bonta2020, Williams2020, Amorim2021, Cackett2021, Dehghanian2021, Bentz2022, Villafana2022, U2022}. Remarkably it has provided a relation between H$\beta$ BLR size and luminosity at 5100 \AA\, ($L_{5100}$) that can be used to estimate BLR size and black hole mass of any AGN having a single-epoch spectrum \citep{Bentz2013}. Reverberation mapping study of high-accreting sources \citep{2016ApJ...820...27D,2016ApJ...825..126D, 2018ApJ...856....6D} show significant deviation from the BLR size-luminosity relation, that is found to be due to Eddington ratio/Fe II strength \citep{Du2019}. Therefore, proper calibration of the size-luminosity relation is crucial by increasing the sample size and objects with diverse properties. Moreover, emission-line and continuum light curves can be used to estimate the BLR size and constrain the geometry and kinematics of the BLR via geometrical and dynamical models \citep[e.g.,][]{10.1093/mnras/stu1419, 10.1093/mnras/staa3828}.

RM of radio-loud AGN is challenging due to the contribution of non-thermal emission, as a consequence RM have been successfully performed only for a few radio-loud AGNs. For example, H$\beta$ lag of 3C120 and 3C273 has been well measured by \citet{Peterson_1998} and \citet{Kaspi2000}, respectively \citep[see also][for 3c273]{2019ApJ...876...49Z}. \citet{2020A&A...642A..59R} successfully measured H$\beta$ lag of PKS1510-089. RM study based on Mg II line of radio-loud AGN has also been carried out successfully for HE 0413-4031 by \citet{2020ApJ...896..146Z}; however, in some cases strong non-thermal contribution prevented reliable lag measurement \citep[e.g.,][]{2019A&A...631A...4N,2020ApJ...891...68C,Amaya-Almazan2022}.

PKS 0736+017 (hereafter PKS0736) is a flat-spectrum radio quasar (FSRQ) located at the redshift of 0.189 \citep{1967ApJ...147..837L,1993MNRAS.263..999T}. The SED of PKS0736  spans from the radio through the $\gamma$-ray wavelengths, consisting of the typical double hump structure of Blazars where the low-energy hump is due to synchrotron emission and high energy one is due to the external Compton process \citep{1988AJ.....95..307I, 1998MNRAS.299..433F, Clements2003, Abdalla2020}. The source shows a one-sided parsec-scale jet \citep{1998AJ....115.1295K,article} and a compact core \citep{Gower1984, 1984A&A...135..289R} and is speculated to have been hosted by an elliptical galaxy \citep{1998MNRAS.295..799W,1999MNRAS.308..377M,2000A&A...357...91F}. The optical-UV spectrum is characterized by broad emission lines and a big blue bump that is associated with thermal emission from the accretion disk \citep{1975ApJ...201...26B, 1986ApJ...300..216M}. The previous estimates of the black hole mass of PKS0736 based on the BLR size-luminosity relation elucidates 10$^{8}$-10$^{8.73}$ M$_{\odot}$ \citep{1999ApJ...526..579W, McLure2001, Marchesini2004a, Dai2007, Abdalla2020}. Therefore, the black hole mass of PKS0736 remains highly uncertain, and accurate measurement of the black hole mass will help SED modeling. Moreover, the BLR of PKS0736 is poorly studied due to the unavailability of long-duration spectroscopic monitoring.

In this paper, we performed a spectroscopic variability study of PKS0736 to estimate the size of the BLR and black hole mass through reverberation mapping. The optical data is taken from the Steward Observatory (SO) as a part of the spectropolarimetric monitoring project \footnote{\url{http://james.as.arizona.edu/~psmith/Fermi/}}, which is a support program for the Fermi Gamma-Ray Space Telescope. Apart from the 3C273 reported by \citet{Zhang2019} and PKS 1510-089 reported by \citet{2020A&A...642A..59R}, PKS0736 have good quality data with more than 100 spectra suitable for reverberation mapping based on H$\beta$ emission line. The optical spectrum shows a blue continuum and the presence of strong Balmer lines (H$\beta$ and H$\gamma$) and Fe II emission. We performed several methods to calculate the size of the BLR and the black hole mass of PKS0736 from time-series analysis. In section \ref{data} we describe the data analysis, and in section \ref{Results}  we present the result of the spectral analysis and time-lag measurements using various methods. Subsequently, in section \ref{Discussion} we discuss the result and conclude in section \ref{Conclusions}.

\section{Data}
\label{data}

\subsection{Optical data}
Spectroscopic monitoring of PKS0736 was performed in SO as a part of spectropolarimetric monitoring project \citep{2009arXiv0912.3621S}, which is a support program for the Fermi Gamma-Ray Space Telescope. Observations were carried out using the 2.3m Bok Telescope on Kitt Peak and the 1.54m Kuiper Telescope on Mount Bigelow in Arizona using the spectrophotometric instrument SPOL \citep{1992ApJ...398L..57S} with a 600mm$^{-1}$ grating, which provides the spectral range of 4000\textendash7550{\AA} with a dispersion of 4 {\AA}/pixel. Depending on the width of the slit used for the observation, the resolution is typically between 16-24 {\AA}. Details regarding observations and data reduction are given in \citet{2009arXiv0912.3621S}. In short, differential photometry using a standard field star was performed to calibrate photometric magnitudes. The instrumental magnitudes of the AGN and the comparison star are determined by using a synthetic Johnson V bandpass for spectroscopic data. A total of 133 V-band photometric observations carried out between November 2014 and May 2018 were used. Spectra were flux-calibrated using the average sensitivity function derived from multiple observations of several spectrophotometric standard stars throughout an observing campaign. Final flux calibrations were performed, re-scaling the spectrum from a given night to match the synthetic V-band photometry of that night \citep[see][]{2009arXiv0912.3621S}. Therefore, a total of 127 photometrically calibrated spectra obtained between November 2014 and May 2018 were downloaded from the SO database and used in this work. Among the 127 spectra, 107 spectra were observed with a slit width of 4.1$^{\prime\prime}$. 

\subsection{$\gamma$-ray and radio data}
The $\gamma$-ray data within the energy range of 100MeV to 300GeV were collected from the publicly available database of the Large Area Telescope (LAT) on board the Fermi Gamma-Ray Space Telescope \citep{2009ApJ...707...55A} between December 2014 and November 2017. The weekly-binned reduced light curve was downloaded and used without any further processing from the light curve Repository of LAT\footnote{\url{https://fermi.gsfc.nasa.gov/ssc/data/access/lat/LightCurveRepository/source.html?source_name=4FGL_J0739.2+0137}} \citep{2021ATel15110....1F}. 

The 15GHz radio data observed using the 40m Telescope at the Owens Valley Radio Observatory (OVRO\footnote{\url{https://sites.astro.caltech.edu/ovroblazars/}}) were also collected \citep{2011ApJS..194...29R}.

\section{Results and Analysis}
\label{Results}
\subsection{Optical spectral analysis and lightcurves}
First, the spectra are brought to the rest frame, then emission line fluxes are measured. Two methods are usually adopted in previous studies for flux measurement (1) direct integration of emission lines within a given window after subtracting the power-law continuum \citep[e.g.,][]{Kaspi2000,  Grier2012, Fausnaugh2017, 2018ApJ...856....6D} and (2) detailed multi-component spectral decomposition \citep[e.g.,][]{Barth2015,2019ApJ...886...93R} which also isolates Fe II emission. However, the latter method demands high S/N spectra, while the former works well even with moderate-low S/N spectra. Both methods are found to provide consistent lag estimates \citep{2016ApJ...820...27D}. Here we adopted the direct integration method due to its better robustness and applicability to the moderate S/N spectra like the one used in this work. 

\begin{figure}
\flushleft
\resizebox{8.5cm}{6cm}{\includegraphics{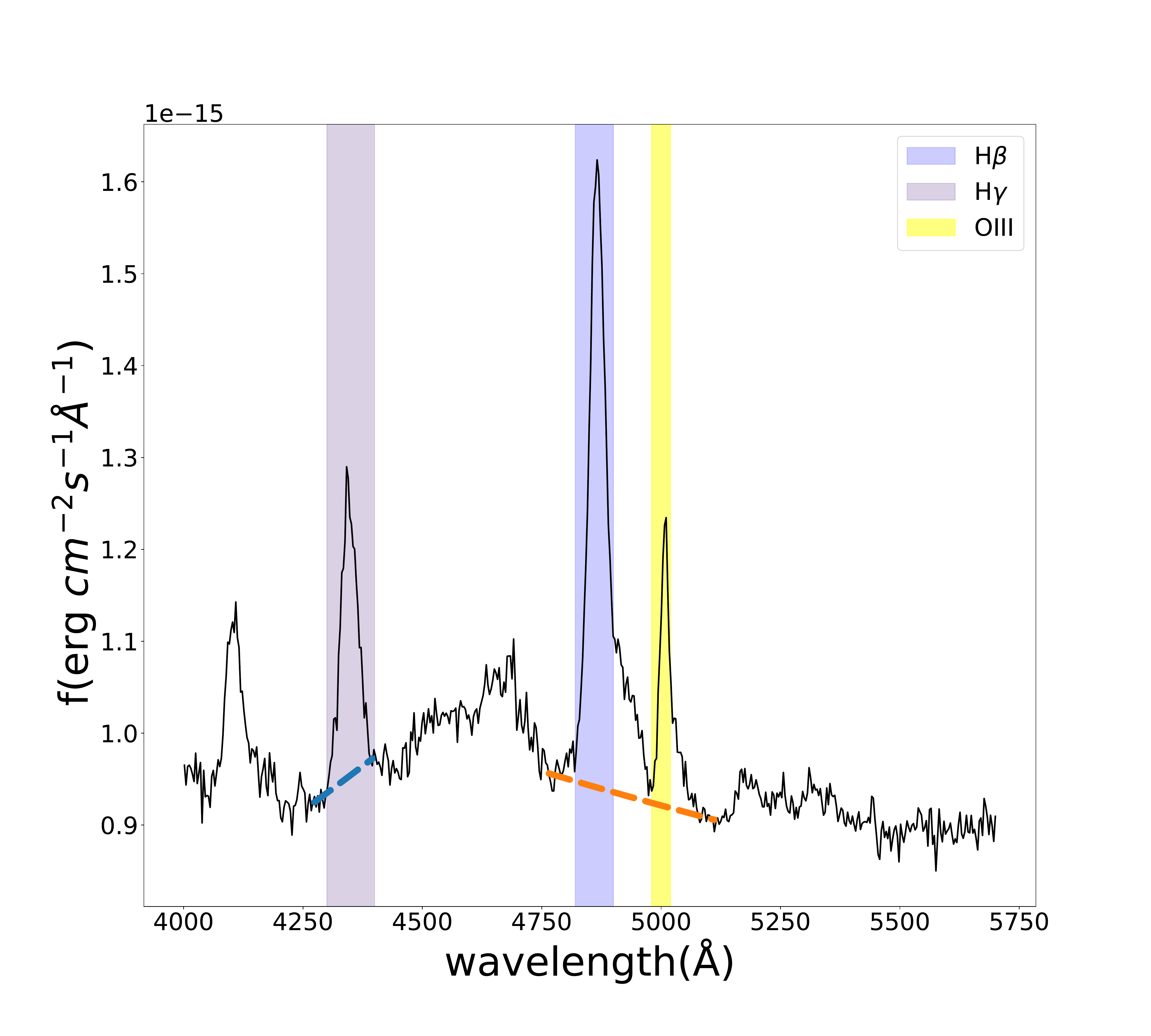}}
\caption{Optical spectrum of PKS0736 on MJD=57046.28, showing the presence of several emission lines. The shaded region represents the line integration windows, and the dashed lines show the best-fit continuum.}
\label{fig:spectrum plot}
\end{figure}

To calculate the emission line flux, we have fitted a power-law i.e., P($\lambda$) = $\alpha \lambda^{\beta}$ where $\alpha$ is a normalization constant, and $\beta$ is the spectral index, locally for H$\beta$ plus [OIII] region in the wavelength window 4760-4790\AA\, and 5080-5120\AA, and separately for H$\gamma$ region in the wavelength window of 4270-4290\AA\, and 4400-4405\AA\, as shown in Fig. \ref{fig:spectrum plot}. Then, we subtracted the best-fit power-law model from each spectrum for both regions separately. Then using the power-law subtracted spectrum, we have integrated the flux values of each emission line region within given wavelength windows as mentioned in Table \ref{tab:region}. It provides us with the flux for each line in each respective spectrum. Note that we consider the region 4820-4900 \AA\ to calculate H$\beta$ line flux because beyond 4900 \AA\, there is an excess contribution from the Fe II 4924 \AA\ multiplets and [OIII]4959 .  This eventually creates a red asymmetry in the line profile of H$\beta$ \citep[see e.g.,][]{1988A&A...192...87J}. Since the Fe II emission is also variable in AGNs it will also be present in the RMS spectrum and hinder the calculation of width of the line emission of H$\beta$.

\begin{figure*}
\resizebox{15cm}{16cm}{\includegraphics{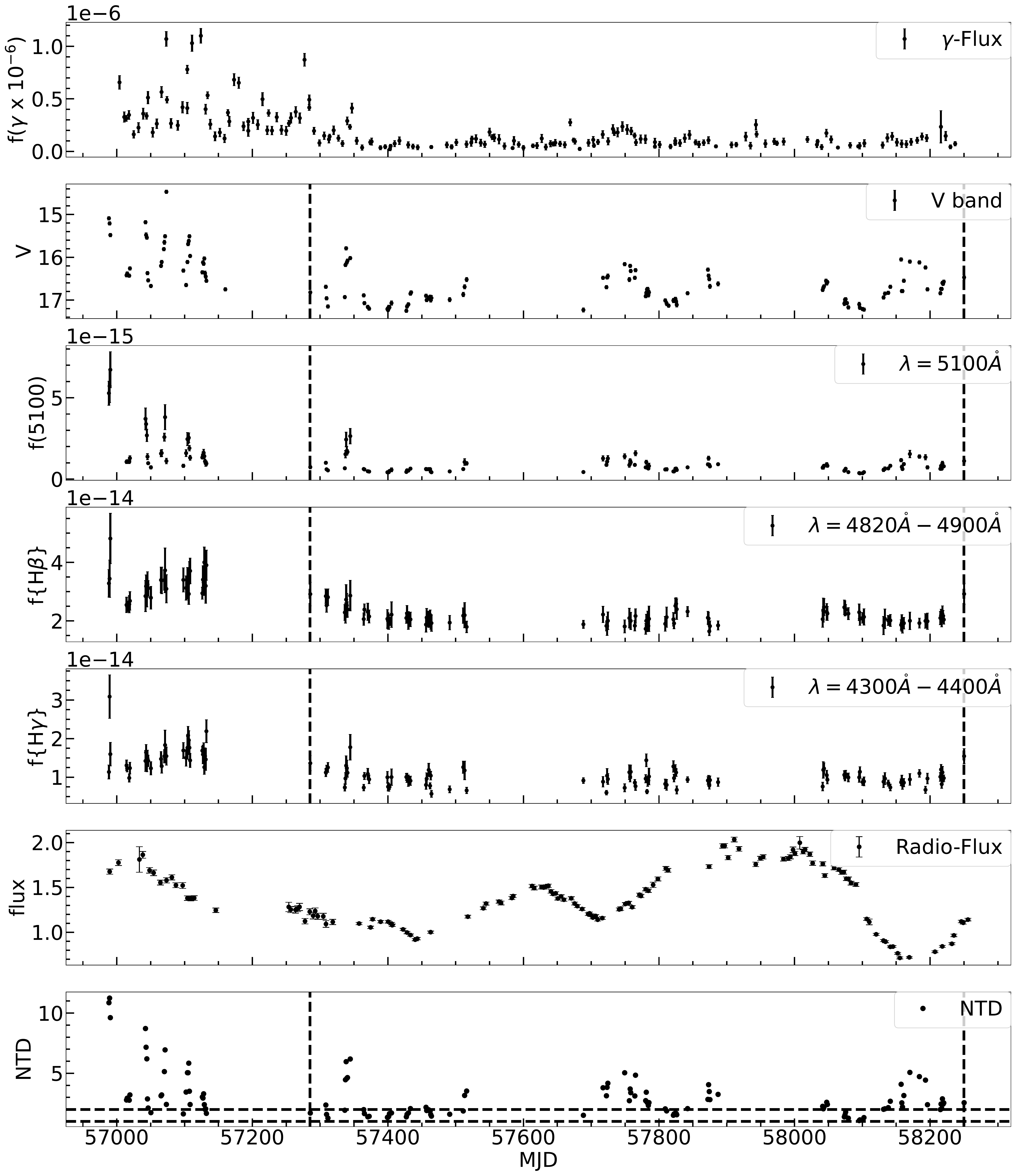}}
\caption{light curves of PKS0736. Top to bottom: $\gamma$ ray flux in photons s$^{-1}$ cm$^{-2}$, photometric V-band magnitude, spectroscopic flux at 5100{\AA}  in ergs cm$^{-2}$s$^{-1}$ \AA$^{-1}$, H$\beta$ and H$\gamma$ flux in ergs cm$^{-2}$s$^{-1}$,  radio flux in Jansky(Jy), Non thermal dominance parameter (NTD). The vertical line at 57285 and 58250 is the MJD range which is used for time analysis and horizontal lines represent NTD=1 and 2.}
\label{fig:lightcurves}
\end{figure*}

We have generated the light curves using the integrated flux values and the photometric V-band light curve obtained from SO. Using the best-fit power-law used for H$\beta$ and the [OIII] region, we have calculated the continuum flux at 5100\AA. As mentioned in the previous section, the flux calibration of SO spectra is performed using standard stars observed throughout the campaign; therefore, seeing effect, slit loss, and systematic uncertainty contributes to the high uncertainty in the flux calibration. Hence, we re-scaled the flux values based on the [O III] $\lambda$5007 emission lines considering it non-variable during the time-scale of the monitoring program. The final light curves are shown in Fig. \ref{fig:lightcurves}. Uncertainties in the calculated emission line flux and continuum flux include systematic uncertainty and Poisson noise added in quadrature. The systematic uncertainties are estimated by the median filter method as described in  \citet{Zhang2019}, first, smoothing the light curve by a median filter with five points and then subtracting the smoothed light curve from the original one. The standard deviation of the residual is used as the estimate of the systematic uncertainty. The flux values are provided in Table \ref{tab:dat points}.

\begin{table}
	\centering
	\caption{Emission line region enlisted to calculate their respective flux.}
	\begin{tabular}{lc} 
		\hline
		Line & Wavelength range (\AA)\\
		\hline
		H$\gamma$& 4300-4400\\
		Fe II    & 4435-4685\\
		H$\beta$ & 4820-4900\\
		OIII     & 4980-5020\\
		\hline
	\end{tabular}
	\label{tab:region}	
\end{table}

\subsection{Variability}
To quantify the variability amplitude of the light curves, we estimated the fractional variability amplitude (F$_{\mathrm{var}}$) for V-band, $f_{5100}$, H$\beta$, and H$\gamma$ light curves using the following equation \citep{RodriguezPascual1997}. 
\begin{align}
F_{\mathrm{var}}= \dfrac{\sqrt{(\sigma^{2}-<\sigma^{2}_{\mathrm{err}}>)}}{<f>}
\end{align}
where $\sigma^{2}$ is the variance, $\sigma^{2}_{\mathrm{err}}$ is the mean square error, and <$f$> is the arithmetic mean of the light curves. The ratio of maximum to minimum flux variation ($R_{\mathrm{max}}$) was also calculated for photometric and spectroscopic light curves. The values are given in Table \ref{tab:flux variation}. All the continuum and line light curves show significant variations with F$_{\mathrm{var}}$=21\% for H$\beta$ and 30\% for H$\gamma$. The $R_{\mathrm{max}}$ is found to 3 and 5 for H$\beta$ and H$\gamma$, respectively. The higher variability found in H$\gamma$ compared to H$\beta$ agrees well with the previous finding in the literature \citep[see][]{Bentz2010,Zhang2019} and as predicated from photo-ionization calculation \citep{2004ApJ...606..749K}. 

Between MJD=57000 - 57285, the $\gamma$- ray flux is in an active or flaring state, and the same is reflected in the optical flux showing correlated variation. It suggests a strong non-thermal contribution from the jet dominates this part of the light curve. Although, the same effect is not visible in the emission line fluxes. Considering the source is FSRQ, a non-thermal contribution from the jet to the optical flux is expected. To determine the dominance of non-thermal contribution over the thermal emission from the accretion disk, we calculated the non-thermal dominance \citep[NTD;][]{Shaw_2012} using 
\newline
\begin{align}
\mathrm{NTD}= \dfrac{L_{o}}{L_{p}} \quad \mathrm{and} \quad L_{o} = L_{\mathrm{disk}}+L_{\mathrm{jet}}
\end{align}
where $L_{o}$ is the observed continuum luminosity having a combination of luminosity emitted from the accretion disk ($L_{\mathrm{disk}}$) and the jet ($L_{\mathrm{jet}}$).  The $L_{p}$ is the predicted disk continuum luminosity that can be assumed to be responsible for line luminosity coming out from the BLR. Therefore, if the thermal emission from the disk only ionizes the broad line clouds and no effect of non-thermal emission is present for the same, then, $L_{p}$ = $L_{\mathrm{disk}}$ and NTD will equalize to 1 + $L_{\mathrm{jet}}$ /$L_{\mathrm{disk}}$. If only thermal contribution from the disk is responsible for continuum luminosity, NTD becomes unity. NTD can be larger than 2 if the jet contribution is greater than the disk contribution.

To measure $L_{p}$, we used the correlation of $L_{\mathrm{H\beta}}$-$L_{5100{\AA}}$ obtained by \citet{2020ApJS..249...17R} for SDSS DR14 quasars and the $L_{\mathrm{H\beta}}$ estimated in this work via $\log L_{\mathrm{H\beta}}=(1.057\pm 0.002) \log L_{5100{\AA}}+(-4.41\pm0.10)$. The variation of NTD with time is shown in the last panel of Fig. \ref{fig:lightcurves}. We note the following points: (1) The NTD varies between $1.1-11.2$ with an average of 2.7, suggesting that the non-thermal emission from the jet contributes to the continuum variation, and (2) From MJD = 57000 - 57285; the NTD shows strong spikes, which are correlated with the flaring event in the $\gamma$-ray light curve, increasing up to NTD=11.

\begin{figure}
\flushleft
\resizebox{8cm}{15cm}{\includegraphics{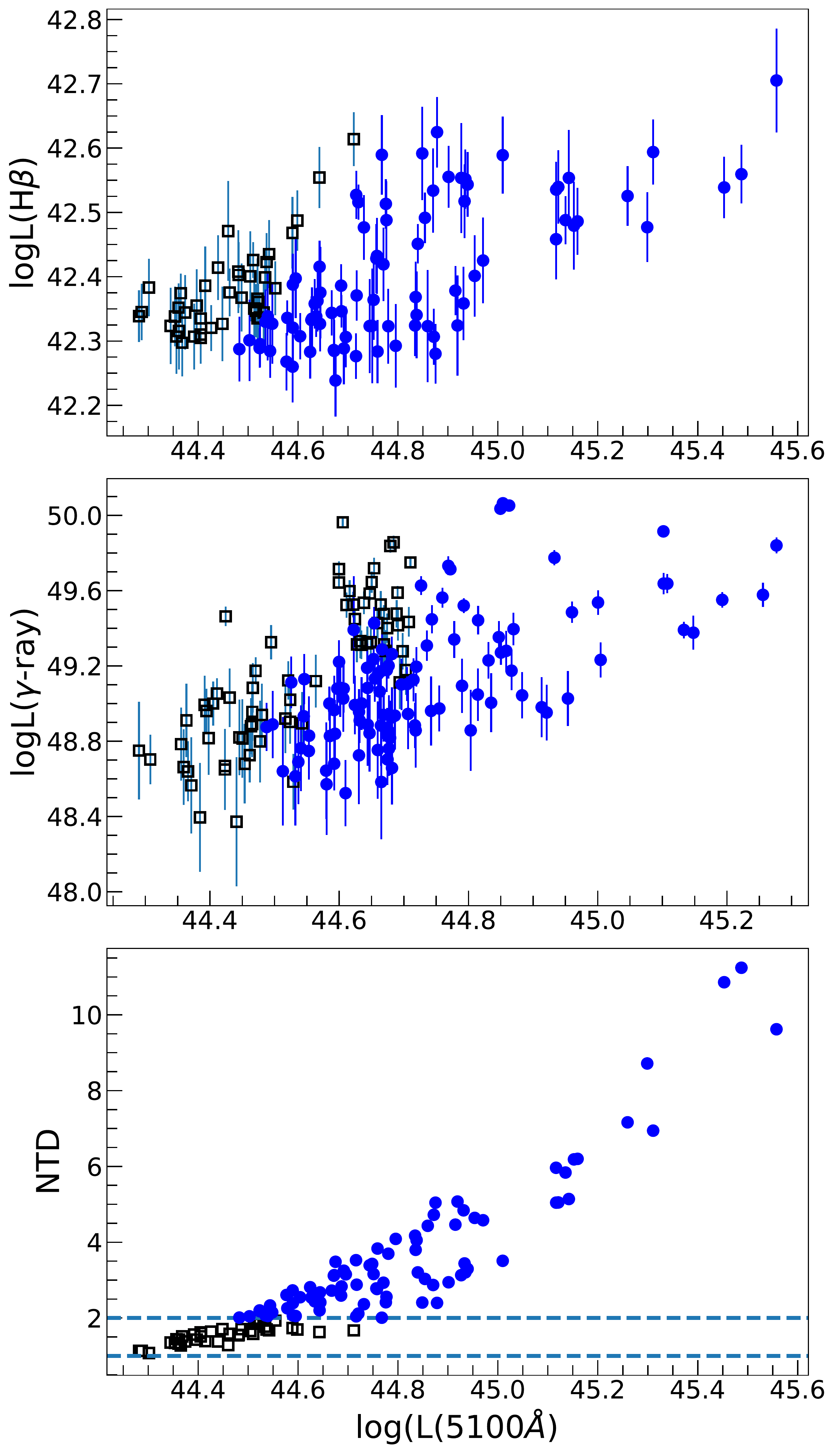}}
\caption{Top to bottom: Plots shown are $\log$ L(H$\beta$), $\log$ L($\gamma$) and Non thermal dominance parameter (NTD) against $\log$ $L_{5100}$. The black empty square represent data points with NTD$<$2 while the filled blue circles are for NTD$\geq$ 2. The dotted horizontal lines are at NTD=1 and 2.}
\label{fig:NTD plot}
\end{figure}

The correlation between the continuum and emission-line luminosity of PKS0736 is studied. In Fig. \ref{fig:NTD plot}, H$\beta$ luminosity (upper panel), $\gamma$-ray (middle), NTD (bottom panel) are plotted against 5100{\AA} continuum luminosity. The NTD is larger than 2 when $\log L_{5100}$ > 44.6. Therefore, a significant non-thermal contribution in the optical emission is found above this luminosity and disk contribution found to dominate below this range. This is also clear from the correlation of line and continuum luminosity plot, which shows a large scatter above $\log L_{5100}$ > 44.6 (see upper panel of Fig. \ref{fig:NTD plot}). The Spearman correlation coefficient calculated for $L_{\mathrm{H\beta}}$ vs. $L_{5100}$ is noted as $r_s=0.61$ with a p-value of no correlation to be $p=10^{-15}$ whereas a much stronger and positive correlation is seen for data points where NTD<2 with $r_s=0.71$ and $p=10^{-7}$. As shown in the Fig.\ref{fig:lightcurves}, the $\gamma$-ray flux changes by a factor of 20 between MJD=57000-57285 with the peak flux of 1.1$\times 10^{-7}$ photons s$^{-1}$ cm$^{-2}$ on MJD=57124.0. Apart from the flaring event during MJD=57000-57285, the light curve is mostly quiescent. The $F_{\mathrm{var}}$ is found to be 99.5\% for full light curve and 51\% for MJD>57285. Therefore, the variations in $\gamma$-ray is much larger than the optical flux variation. Optical continuum flux is found to be correlated with the $\gamma$-ray flux with $r_s=0.79$ and $p=10^{-15}$ (see Fig.\ref{fig:NTD plot}).  

\begin{table*}
	\centering
	\begin{tabular}{lcccc} 
		\hline
		MJD & V & f(5100) & f(H$\beta$) & f(H$\gamma$)\\
		(1) & (2) &(3) &(4) & (5) \\
		\hline
		 56989 & 15.09 & 5.28$\pm$0.59 & 29.15$\pm$3.20 & 13.63 $\pm$1.62 \\
         56990 & 15.21 & 5.71$\pm$0.49 & 28.45$\pm$4.14 & 11.22 $\pm$1.63 \\
         56991 & 15.48 & 6.72$\pm$1.18 & 25.32$\pm$2.64 & 15.98 $\pm$2.95 \\
         57015 & 16.42 & 1.06$\pm$0.12 & 28.09$\pm$4.19 & 13.05 $\pm$1.55 \\
		\hline
	\end{tabular}
	
	\caption{Columns as: (1) Modified Julian date (MJD), (2) photometric V-band magnitude, (3) spectroscopic flux at 5100$\AA$ in units of 10$^{-15}$ ergs cm$^{-2}$s$^{-1}$ \AA$^{-1}$, (4) and (5) are H$\beta$ and H$\gamma$ line flux in units of 10$^{-15}$ ergs cm$^{-2}$s$^{-1}$,  respectively. This table is available in its entirety in machine-readable form. A portion is shown here for guidance. }
	\label{tab:dat points}
\end{table*}

\begin{table}
	\centering
	\caption{Variability statistics.}
	\begin{tabular}{lccc} 
		\hline
		light curve & Median flux & $F_{\mathrm{var}}$ ($\%$) & $R_{\mathrm{max}}$\\
		(1) & (2) & (3) & (4)\\
		\hline
		V-band     &0.76$\pm$0.07&69.86$\pm$4.30&12.94$\pm$0.67\\
		f$_{5100}$ &0.82$\pm$0.11&85.43$\pm$5.57&18.87$\pm$4.11\\
		H$\beta$   &21.95$\pm$0.66&21.10$\pm$1.79&2.92$\pm$0.63\\
		H$\gamma$  &10.11$\pm$0.41&29.73$\pm$2.25&5.41$\pm$0.88\\
		\hline
	\end{tabular}
	\label{tab:flux variation}
\begin{tablenotes}
\item \textbf{Notes.} Columns are: (1) light curve and (2)  median flux of the light curve in units of 10$^{-15}$ ergs s$^{-1}$cm$^{-2}$ \AA$^{-1}$ for $f_{5100}$, and 10$^{-15}$ ergs s$^{-1}$cm$^{-2}$ for emission lines; (3) fractional rms variability in percentage; and (4) the ratio of maximum to minimum flux variation.
\end{tablenotes}
\end{table}

\subsection{Time delay measurement}
As mentioned in the previous section, a strong flare in the $\gamma$-ray light curves is seen between 57000-57285 MJD due to the dominance of non-thermal contribution over the disk contribution. Therefore, we excluded this region and used the time window between 57285 and 58250 for further analysis.    

\subsubsection{Interpolated cross Correlation Analysis}
The cross-correlation technique quantifies the degree of similarity between two sets of numbers. The process involves two light curves to assess information between peak/centroid values. One can shift either continuum concerning emission-line light curve or vice versa and calculate the correlation coefficient ($r$) at each shifted time. Then the peak or the centroid ($\tau_{\mathrm{cent}}$) covering 0.8$\times r_{\mathrm{peak}}$ is considered as the lag \citep{Peterson_2004}. 
Following \citet{Peterson_1998}, we performed interpolated cross-correlation (ICCF) \footnote{\url{https://bitbucket.org/cgrier/python_ccf_code/src/master/}}. First, the cross-correlation function (CCF) is measured by interpolating the continuum series keeping the emission-line light curve unchanged, and second interpolating the emission-line light curve keeping the continuum light curve unchanged. The final ICCF is determined by averaging these two results. The CCF is characterized by (1) its peak value $r_{\mathrm{max}}$, (2) the time delay corresponding to this value $\tau_{\mathrm{peak}}$, and (3) the centroid $\tau_{\mathrm{cent}}$ of the CCF. The Monte Carlo method of flux randomization (FR) and random subset selection (RSS) is used to estimate the uncertainty in the lag  \citep{Peterson_1998, Peterson_2004}.

\begin{figure*}
\begin{minipage}[b]{0.33\textwidth}
\includegraphics[scale=0.55,width=1\textwidth]{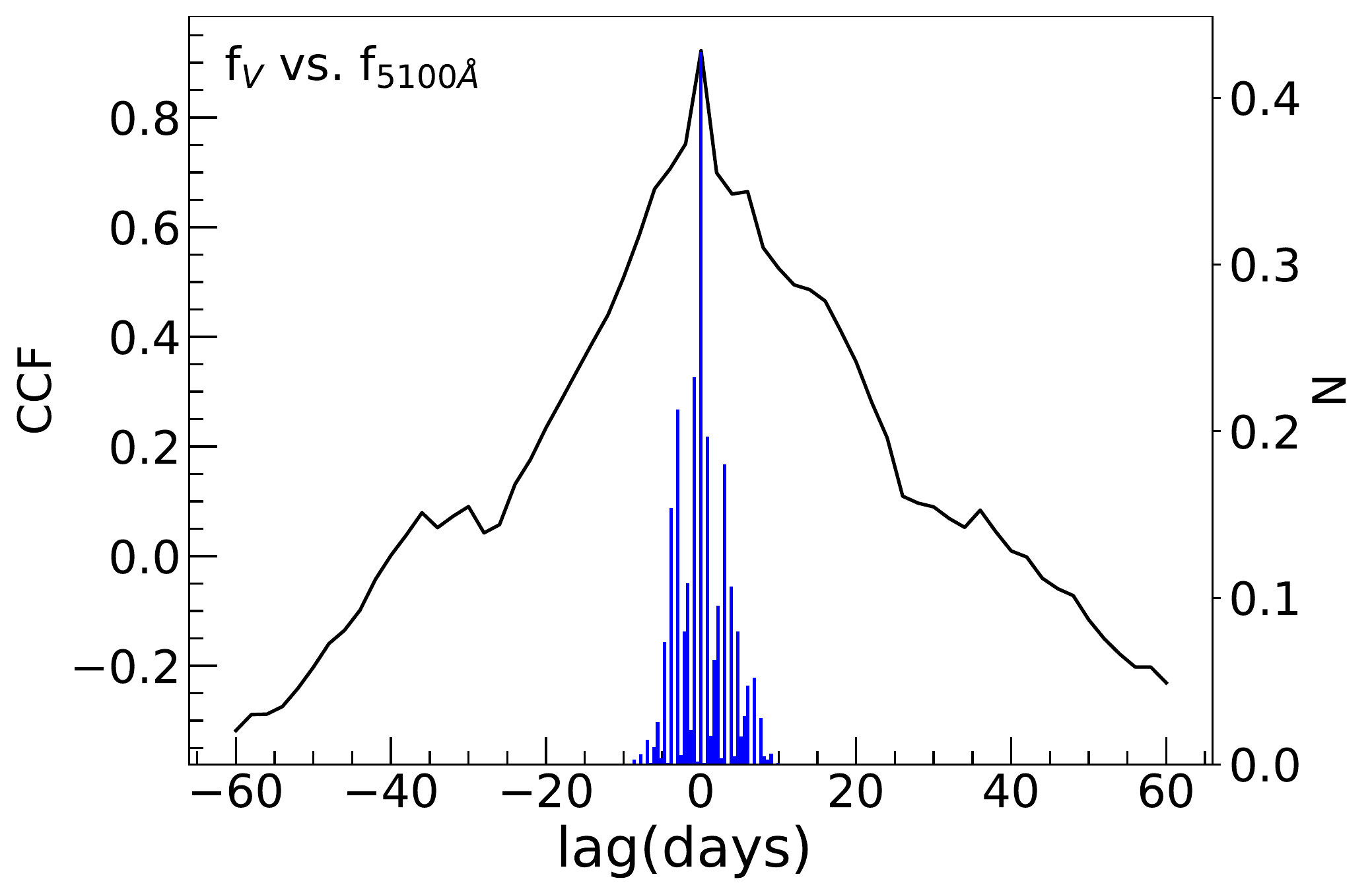}
 \end{minipage}%
\begin{minipage}[b]{0.33\textwidth}
\includegraphics[scale=0.55,width=1\textwidth]{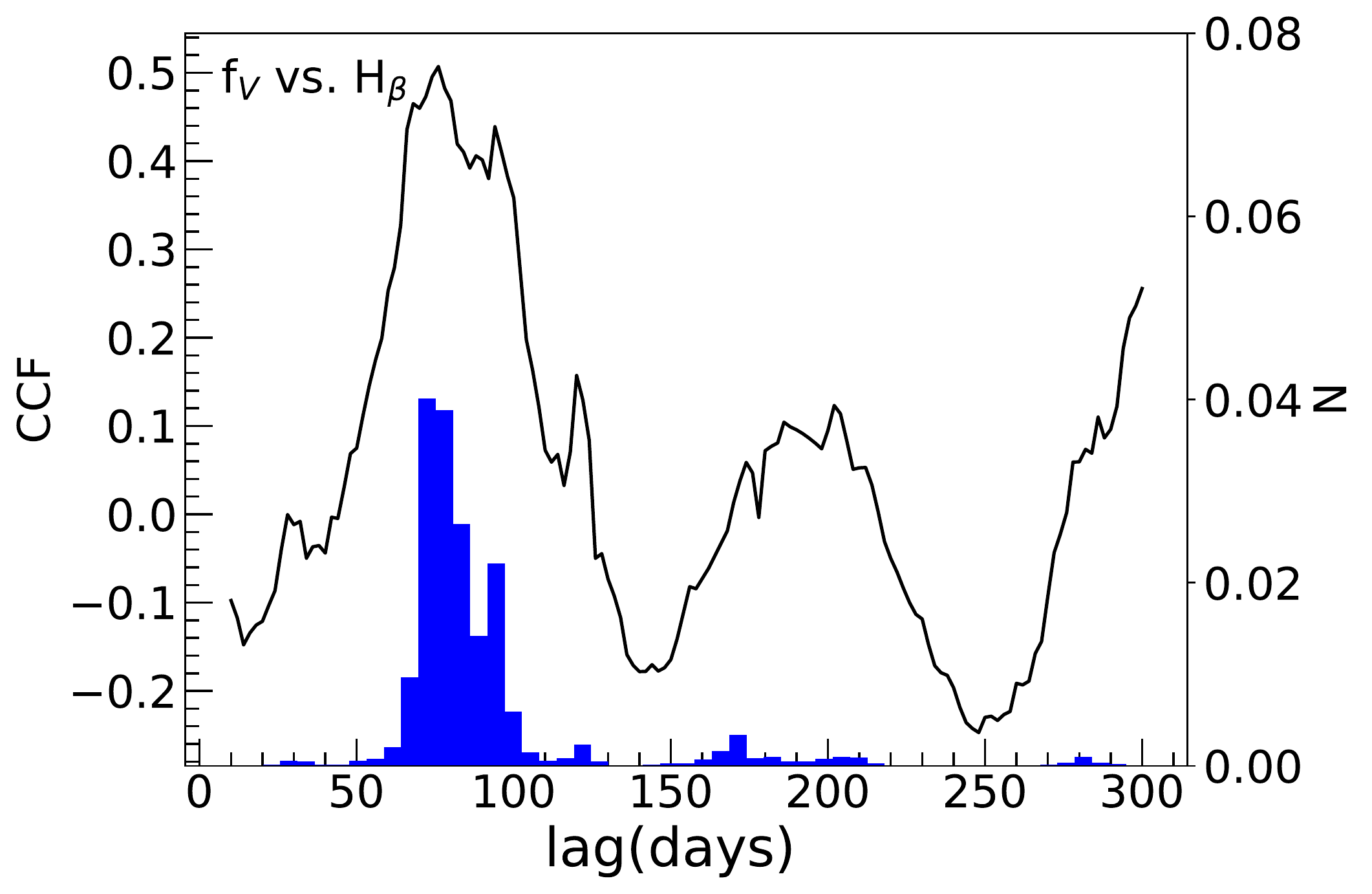}
\end{minipage}
\begin{minipage}[b]{0.33\textwidth}
\includegraphics[scale=0.55,width=1\textwidth]{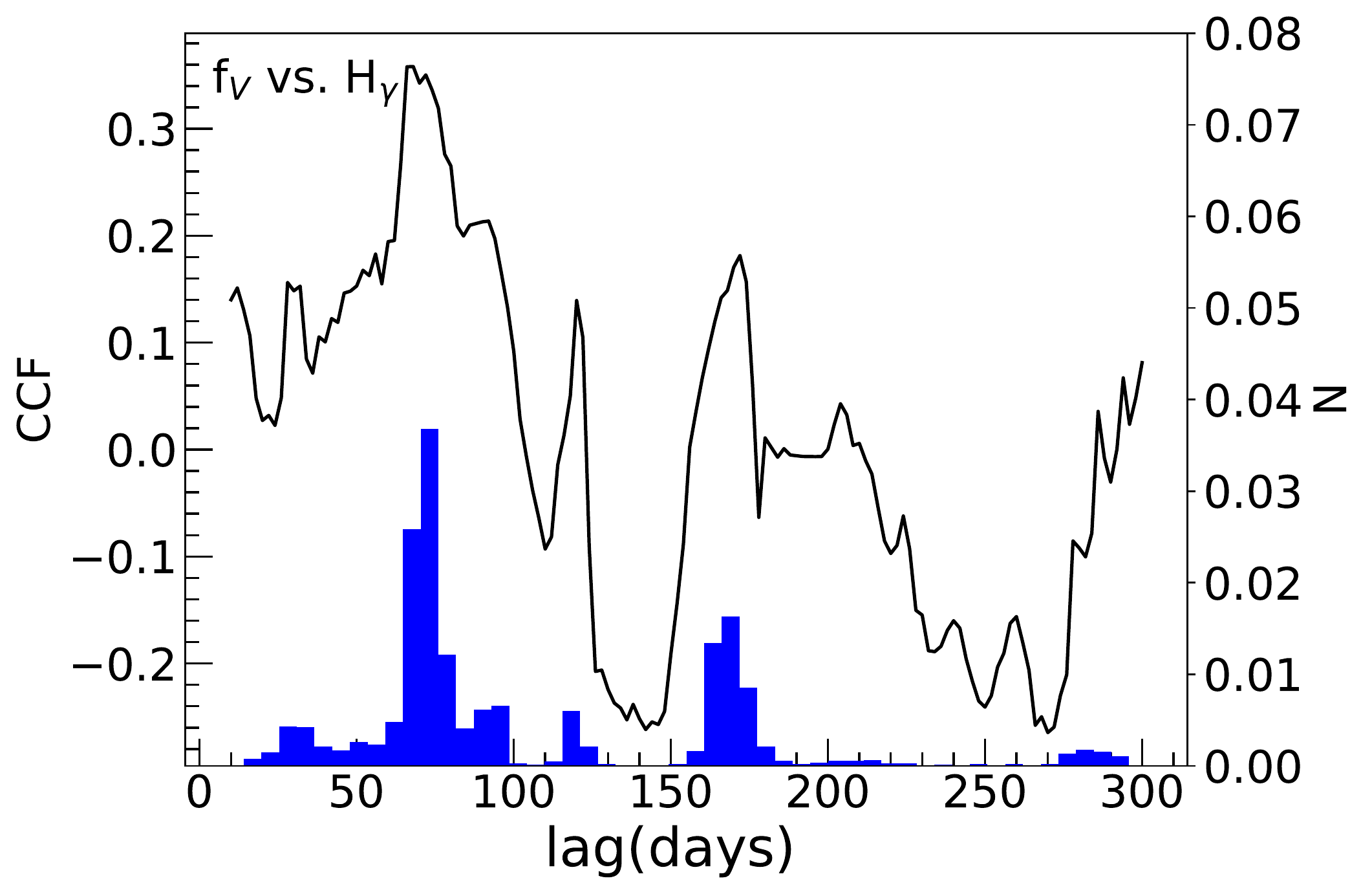}
\end{minipage}
\caption{Cross-correlation function and lag distribution for V-band vs $f_{5100}$ (left), H$\beta$ (middle) and H$\gamma$ (right). The ICCF (line) and centroid probability distribution from ICCF (filled histogram) are shown.}
\label{fig:CCF plots}
\end{figure*}

We have performed the auto-correlation (ACF) between the two continuum light curves, the 5100{\AA} and photometric V-band in Fig. \ref{fig:CCF plots}. The ACF shows a lag of zero-day as expected. We also estimated the time lag between photometric V-band flux vs. H$\beta$ and H$\gamma$.  
ICCF shows a strong peak around 80 days for H$\beta$ and H$\gamma$, and a minor peak at $\sim$180 days that could be due to the seasonal gaps (see discussion in section \ref{sec:seasonal_gaps}). Therefore, we restricted our analysis to 150 days covering the major peak and estimated lag within that. The ICCF method provides lags between V-band vs. H$\beta$ of  79.0$^{+7.1}_{-5.0}$ days and between V-band and H$\gamma$ of 71.8$^{+10.9}_{-18.1}$ days in the observer-frame. We obtained a shorter lag for H$\gamma$ compared to H$\beta$, which agrees with the previous studies on different objects \citep[e.g.,][]{Bentz2010}.

\subsubsection{JAVELIN}
We have also used {\small JAVELIN}\footnote{\url{https://github.com/nye17/JAVELIN}} developed by \citet{2011ApJ...735...80Z, 2013ApJ...765..106Z} which estimates lag by modeling the continuum and line light curves. According to \citet{2009ApJ...698..895K}, quasar variability can be well described by a damped random walk (DRW) process. {\small JAVELIN} models the continuum light curve using a DRW having two parameters: amplitude and time scale of variability. Then the emission line light curve is generated, which is a shifted, scaled, and smoothed version of the continuum light curve. It interpolates between data points and self-consistently estimates and includes the uncertainties in the interpolation. Error estimation uses the Markov Chain Monte Carlo (MCMC) approach to calculate the statistical confidence limits on each best-fit parameter and accurately model the continuum and emission-line light curve simultaneously.

\begin{figure}
\begin{minipage}[b]{0.25\textwidth}
\includegraphics[scale=0.35,width=1\textwidth]{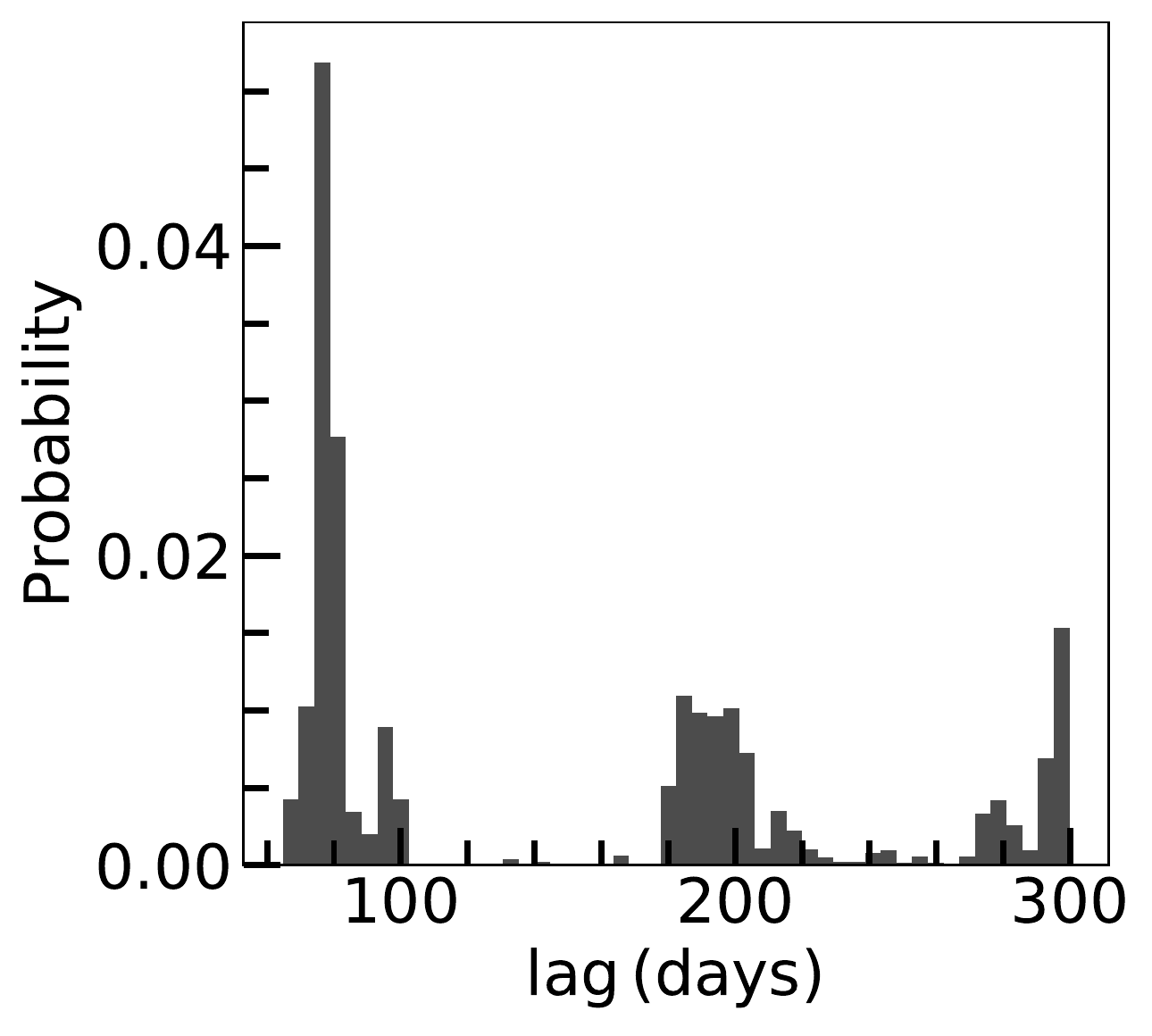}
 \end{minipage}%
\begin{minipage}[b]{0.26\textwidth}
\includegraphics[scale=0.35,width=1\textwidth]{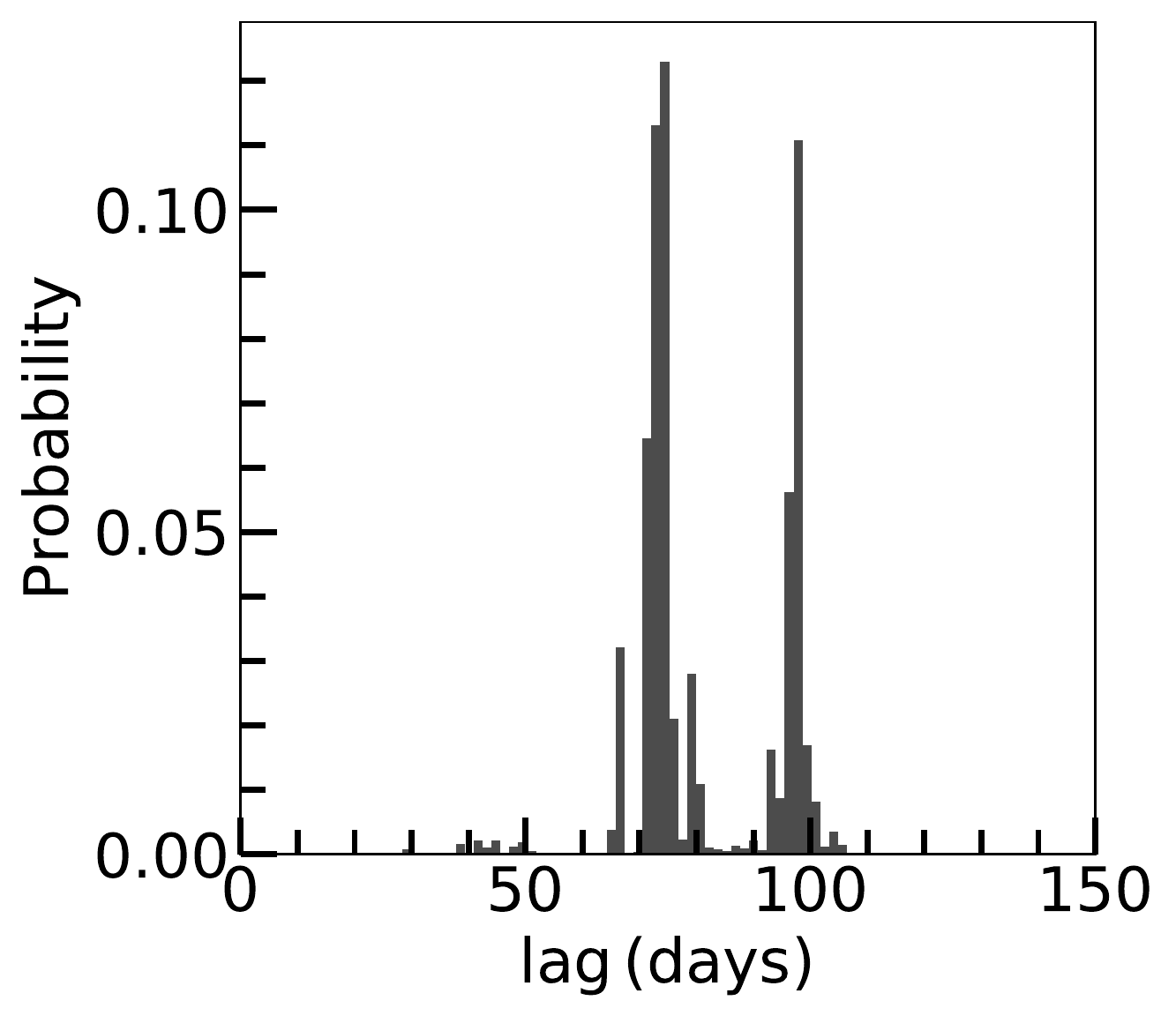}
\end{minipage}
\caption{Probability distribution of observed frame lag computed by {\small JAVELIN} between V-band vs H$\beta$ (left),  and with H$\gamma$ (right).}
\label{fig:JAVELIN plots}
\end{figure}

In Fig. \ref{fig:JAVELIN plots}, the probability distribution of the observed frame lag is plotted, as computed by {\small JAVELIN}. The left panel of the Figure shows the probability distribution of {\small JAVELIN} lag between the V-band light curve vs. H$\beta$; here, too, we noticed a significant peak at $\sim$80 days and multiple minor peaks. Therefore, we calculated a lag between 0 to 150 days, similar to ICCF. We also plotted the probability distribution of lag between the V-band light curve vs. H$\gamma$ in the right panel for the lag range of 0 to 150 days. The plots show prominent peaks at $\sim$ 78 and 75 days for H$\beta$ and H$\gamma$, respectively. These results are consistent with the ICCF. Time delays obtained from {\small JAVELIN} are given in Table \ref{tab:lag table}.

\subsubsection{The von Neumann and Bartels estimators}
This method is based on a mean-square successive-difference estimator \citep[see][]{10.1214/aoms/1177731677} which handles sparsely and irregularly sampled data and is relevant for the process underlying the variability that cannot be adequately modeled. It does not rely on data interpolation, binning in correlation space, ergodicity arguments, and stochastic models for quasar variability. Hence time-delay measurements are restricted to the information embedded in the light curves being processed \citep{Chelouche2017}. 

\begin{figure}
\resizebox{7.8cm}{3.8cm}{\includegraphics{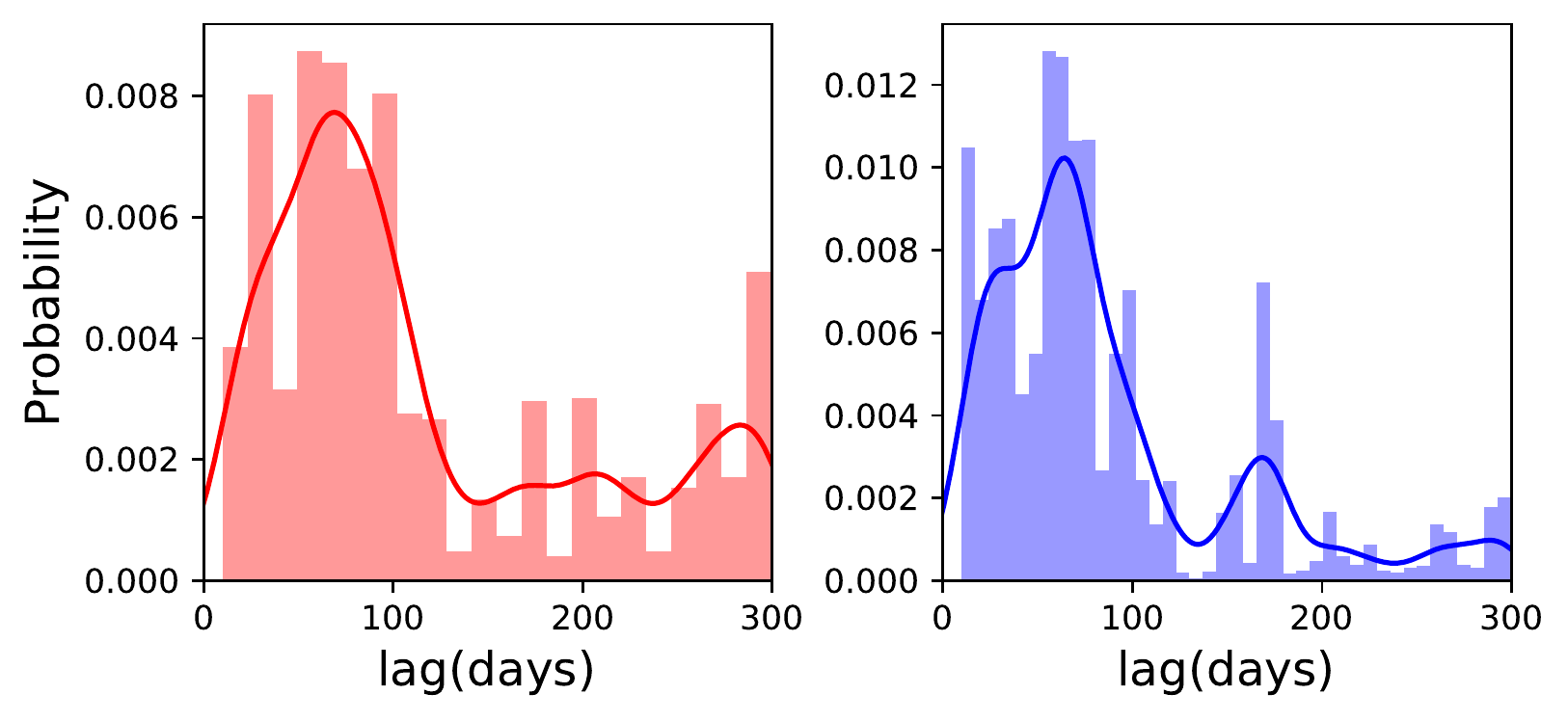}}
\resizebox{8cm}{3.8cm}{\includegraphics{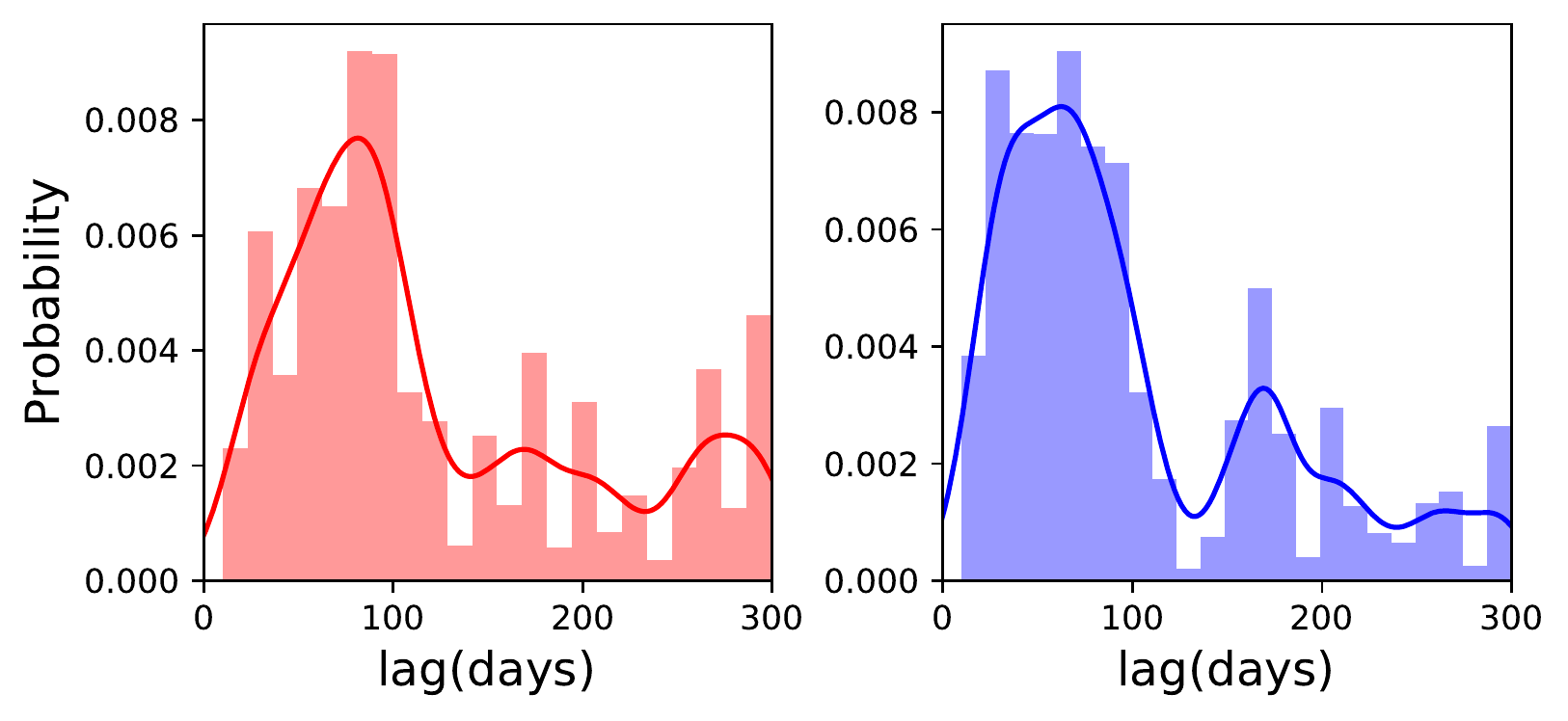}}
\caption{Probability distribution of the observed frame time lag based on the von Neumann estimator (top panels) and the Bartels estimator (bottom panels) for H$\beta$ (left) and H$\gamma$ (right).}
\label{fig:vnrm_plot}
\end{figure}

The time delay distribution obtained from the von Neumann method after a Monte Carlo simulation of FR-RSS, as done for the CCF analysis, is shown in the upper panels of Fig. \ref{fig:vnrm_plot}. The H$\beta$ and H$\gamma$ show firm peaks at $\sim$68 and 60 days, respectively. The Bartels estimator is the modified method of the von Neumann estimator \citep{doi:10.1080/01621459.1982.10477764} which uses a ranked light curve instead of a sorted one, which can also be used to measure time delay based on the regularity or randomness of data. The lag results are given in Table \ref{tab:lag table} between time interval taken from 0-300 days. The results obtained using von Neumann and Bartels methods are consistent with the lags obtained using ICCF and JAVELIN.
\begin{table}
	\centering
	
	\begin{tabular}{lcc} 
		\hline
		Method & Lag (days) &\\
		\hline
		 &V vs. H$\beta$ & V vs. H$\gamma$\\
		 \hline
		ICCF & 79.0$^{+7.1}_{-5.0}$  &71.8$^{+10.9}_{-18.1}$ \\
		JAVELIN & 78.2$^{+6.8}_{-3.5}$ & 74.7$^{+22.7}_{-2.8}$ \\
		von Neumann & 67.6$^{+37.6}_{-31.4}$ & 60.0$^{+33.5}_{-31.6}$\\
		Bartels & 73.7$^{+37.5}_{-28.3}$ & 62.2$^{+31.4}_{-31.1}$ \\
		\hline
	\end{tabular}
	\caption{Columns as follows: (1) method used to calculate the lag(days), (2) depicts lag in observed frame between photometric V-Band vs. H$\beta$ and (3) lag in observed frame between photometric V-Band vs. H$\gamma$ }
	\label{tab:lag table}
\end{table}

\begin{figure}
\centering
\resizebox{6.5cm}{5cm}{\includegraphics{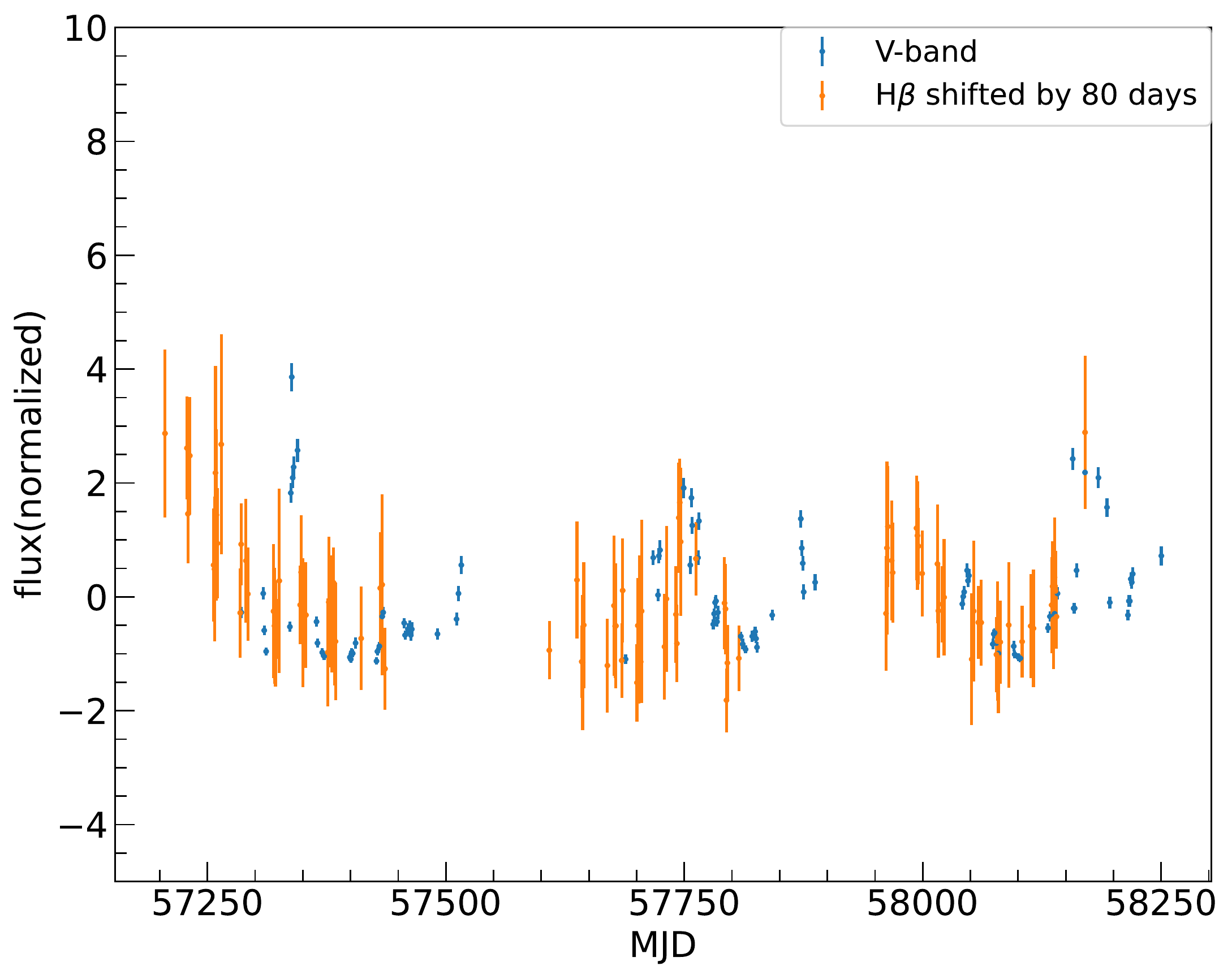}}
\caption{Normalized V-band light curve plotted along with the H$\beta$ light curve back-shifted by 80 days.}
\label{fig:shifted lightcurve plot}
\end{figure}

The above methods suggest a time lag of $\sim$80 days between the photometric V-band continuum and the H$\beta$ line light curve. We visually analyze the consistency of the measured lag; the V-band flux light curve along with the H$\beta$ light curve back-shifted by 80 days are plotted in Fig. \ref{fig:shifted lightcurve plot}. The continuum and back-shifted line light curves agreeably match. Therefore, we considered the lag of 79.0$^{+7.1}_{-5.0}$ days in the observed frame obtained by the ICCF method as the best lag measurement for PKS 0736+017.

\subsubsection{Mean and RMS spectra}
We constructed mean spectrum and root-mean-square (rms) spectrum as follows.
\begin{align}
\bar{F(\lambda)} =\dfrac{1}{N}\sum_{i=0}^{N-1}F_{i}(\lambda),
\end{align} 
and
\begin{align}
S(\lambda) =\sqrt{\Big[\dfrac{1}{N-1}\sum_{i=0}^{N-1}(F_{i}(\lambda)- \bar{F(\lambda)}^{2})\Big]}
\end{align}
where F$_{i}(\lambda)$ is the ith spectrum of the N(127) spectra that comprise the database. 
The mean spectrum is the average spectrum of all the spectra obtained during the campaign, whereas the rms spectrum is based on the variation around this mean. 
The rms spectrum isolates the constant features such as narrow emission line and the host galaxy contribution of the spectrum and hinders accurate line width measurements. However, the rms spectrum is often too noisy therefore measuring the line width from the rms spectrum is challenging.   

\begin{figure}
\resizebox{8cm}{8cm}{\includegraphics{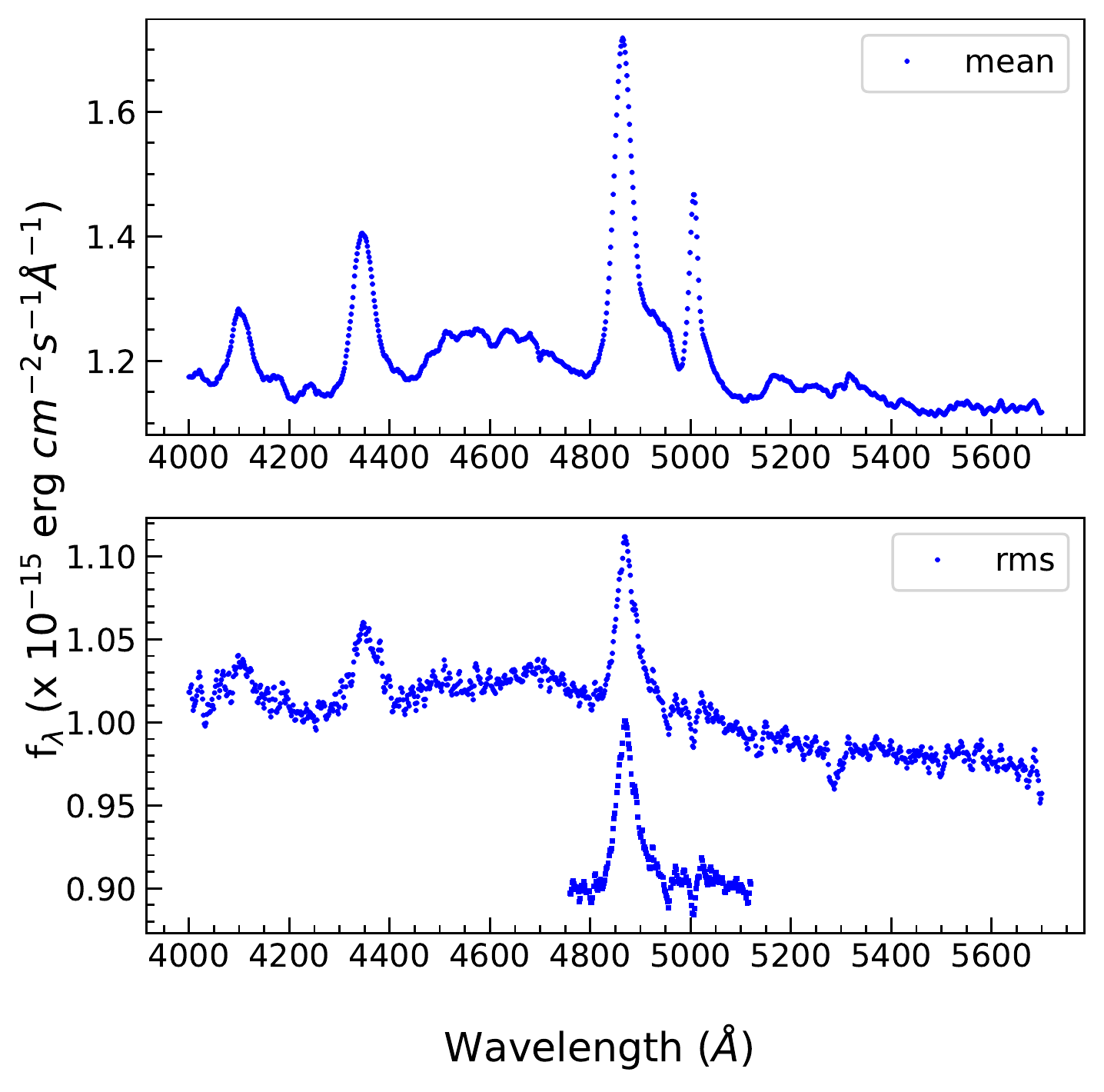}}
\caption{Mean and RMS spectra of PKS0736 constructed from the nightly spectrum after rescaling based on [O III]$\lambda$5007. Top: mean spectrum. Bottom: RMS spectra before (upper plot) and after (lower plot) the continuum subtraction in the region of wavelength window 4760-4790\AA\, and 5080-5120\AA.}
\label{fig:Mean and RMS spectra}
\end{figure}

In Fig. \ref{fig:Mean and RMS spectra}, we showed the mean and rms spectra constructed from all the flux-calibrated spectrum without subtracting from the fitted continuum. The lower panel depicting the rms spectra with and without subtraction from fitted power-law continuum. Both the mean and rms spectra show the presence of strong Balmer lines. Most importantly, the narrow [O III]5007, present in the mean spectrum, disappeared in the rms spectrum suggesting the flux calibration is proper. As noted earlier, although the SO monitoring campaign has used different slits throughout the campaign, the majority of them (107 out of 127) are obtained using 4.1$^{\prime\prime}$ slit width. To find out any effect of slit width in the construction of the mean and rms spectra and subsequent analysis, we re-constructed them using only spectra taken with 4.1$^{\prime\prime}$ slit. The resultant mean and rms spectra (see Fig. \ref{sec:4arsec_mean_rms}) is remarkably similar to that obtained using the entire spectral data.  

\subsubsection{Detection significance of cross-correlation lag}
Recent reverberation mapping studies on large samples perform simulation to access the lag significance \citep{2022MNRAS.509.4008P,U2022,2022ApJ...926..225H}, however, there are no universal criteria to define the lag reliability. We used the publicly available PyIICCF code\footnote{\url{https://github.com/legolason/PyIICCF/}} developed by  \citet{2022arXiv220706432G} which uses the method described in \citet{U2022} to assess the lag significance of our measurement. In this method, two uncorrelated red-noise light curves are simulated with the same S/N and cadence as the observed data to determine the probability of finding the peak of correlation ($r_{\mathrm{max}}$) larger than the observed ICCF peak via null hypothesis test.  

\begin{figure}
    \centering
    \includegraphics[width=9cm, height=4cm]{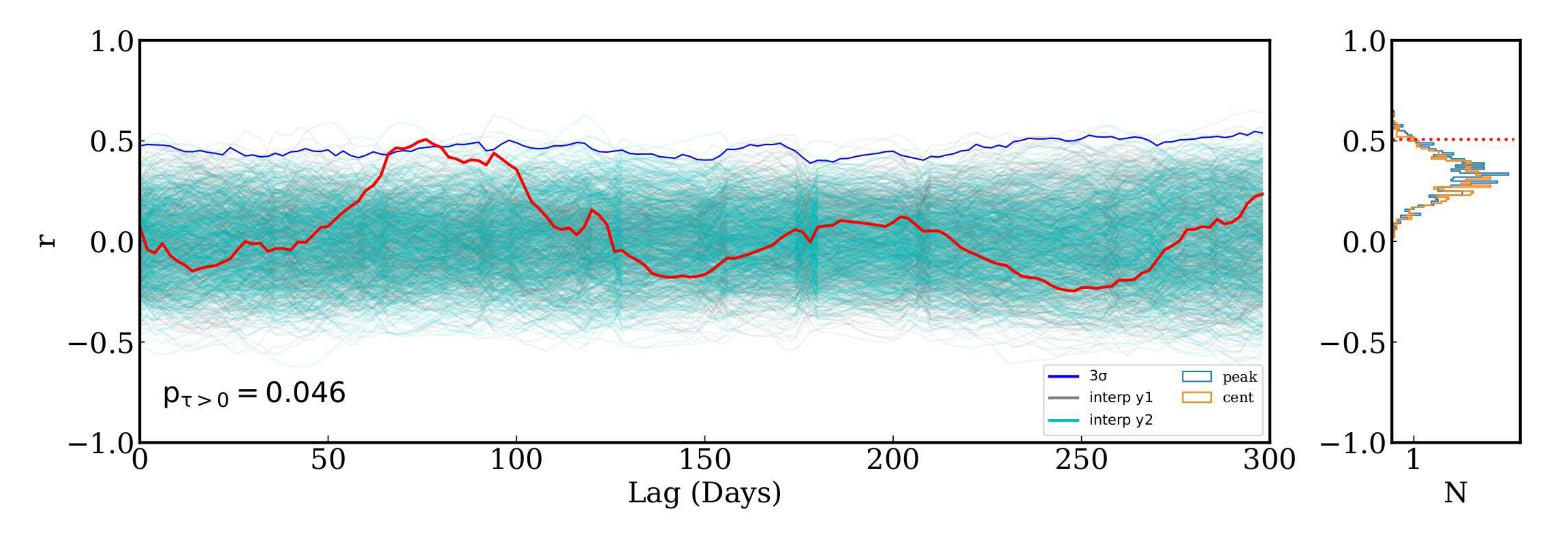}
    \caption{ Left: The plot shows the CCF coefficient(r) Vs. lag between V-band continuum and H$\beta$ line light curve taken in 0-300 days,  for all 1000 simulations and the red curve shows the r Vs. lag for the observed light curves with 3-$\sigma$ confidence interval. Right: CCF peak and centroid distribution of the coefficient and the no of simulations showing r>0.5. }
    \label{fig:significance plot}
\end{figure}

For this purpose, we start by simulating the continuum and line light curves using the DRW model. First, we fitted the continuum V-band and H$\beta$ line light curves using the DRW model and used the best-fit parameters to simulate the corresponding light curves. Second, we simulated a 100 times longer light curve than the observed data and randomly selected a portion of the light curve. Third, we added Gaussian noise to the mock light curve based on the uncertainty in the observed light curve. Forth, we down-sampled the mock light curve to have the same cadence and length as the observed light curve. We simulated 1000 such sets of continuum and line light curves. Finally, we cross-correlated mock continuum light curve with the observed line light curve and vice versa within a lag search range of 0-300 days. From these CCF, we obtained a distribution of $\tau_{\mathrm{cent}}$ and $r_{\mathrm{max}}$ (see Fig. \ref{fig:significance plot}). Then we derived the p($r_{\mathrm{max}}$) which is the ratio of the number of positive lag with $r_{\mathrm{max}}$ larger than the observed $r_{\mathrm{max}}$ and $\tau_{\mathrm{cent}}>0$ to the total number of positive lag. This p($r_{\mathrm{max}}$) provides a way to access the robustness of our lag estimation, a smaller p($r_{\mathrm{max}}$) gives more robust and reliable lag detection. We found p($r_{\mathrm{max}}$) is 0.05 which is $<0.2$ adopted by \citet{U2022} as the limiting condition for reliable lag determination. This suggests that our lag measurement is robust and reliable.

\subsubsection{Line width and black hole mass measurement}
To estimate the black hole masses, we measured the FWHM and the line dispersion ($\sigma_{\mathrm{line}}$) of the H$\beta$ emission line from both the mean and rms spectra after fitting the continuum and subtracting  with the same. The full width at half maximum (FWHM) is calculated by finding 0.5$\times$F$(\lambda)_{\mathrm{max}}$ both from the left ($\lambda_l$) and right side ($\lambda_r$) of the curve and subtracting the two wavelengths i.e, $\lambda_r- \lambda_l$  \citep{Peterson_2004}. Whereas, for calculation of $\sigma_{\mathrm{line}}$, the flux weighted line center was first determined as follows:
\begin{align}
\lambda_{0} =\dfrac{\int \lambda f_{\lambda}\,d\lambda}{\int f_{\lambda}\,d\lambda}
\end{align}
and the line dispersion is
\begin{align}
\sigma_{\mathrm{line}}^{2} =\dfrac{\int \lambda^{2} f_{\lambda}\,d\lambda}{\int f_{\lambda}\,d\lambda} - \lambda_{0}^{2}
\end{align}
The values are mentioned in Table \ref{tab:Black-hole mass table}. The same for spectra taken with 4.1$^{\prime\prime}$ slit is given in Table \ref{tab:Black-hole mass table_B}, which shows consistent results justifying the line width measurements from the entire spectral data. 

Assuming that the black hole’s gravitational potential governs the gas motion in the BLR, black hole mass can be estimated by combining the size of the BLR ($R_{\mathrm{BLR}}$) and the velocity width of broad emission lines ($\Delta$V) based on the virial relation:
\begin{align}
M_{\mathrm{BH}} =  \dfrac{f \times R_{\mathrm{BLR}} (\Delta V)^{2}}{G}
\label{eq:virial eqn}
\end{align}
\newline
where $f$ is a dimensionless scale factor that depends on the kinematics and geometry of BLR gas clouds. The BLR size (calculated from time lag $R_{\mathrm{BLR}}$=c$\tau$) and line width (from mean and rms spectra) with $f$ of 4.47($\sigma$) or 1.12(FWHM) taken from \citep{Woo2015} is used to calculate black hole mass. The Monte Carlo bootstrap method \citep{Peterson_2004} was used to estimate uncertainty in the line width measurements and hence the black hole mass. N spectra were chosen randomly from a set of N spectra without replacement for each realization, and line width was measured from the mean and rms spectra. We varied the endpoints randomly within $\pm$10{\AA} of the initially selected H$\beta$ region (i.e., 4800-4920{\AA}). Finally, a total of 5000 realizations were performed, providing a distribution of FWHM and $\sigma_{\mathrm{line}}$. The mean of the distribution is the final line width, and the standard deviation of the distribution ({ 34 percentile of both sides of the mean i.e. the 16th
and the 84th percentiles of the distribution) was considered the measurement uncertainty. The resultant line width measurement along with black hole mass is given in Table \ref{tab:Black-hole mass table}.

Using the cross-correlation method, we obtained a lag ($\tau_{\mathrm{cent}}$) between optical V-band and H$\beta$ (H$\gamma$) of 66.4$^{+6.0}_{-4.2}$ (60.4$^{+9.2}_{-15.3}$) days in rest frame. We used the virial relation with $f$ of 4.47(1.12) for $\sigma_{\mathrm{line}}$(FWHM) providing four different black hole masses based on the four different choices of line widths. We note that $\sigma_{\mathrm{line}}$ is less sensitive to the line peak and that FWHM is less sensitive to the line wing; therefore, black hole masses based on the $\sigma_{\mathrm{line}}$ are widely adopted as the best mass measurement \citep[e.g.,][]{Peterson_2004,2014SSRv..183..253P}. Therefore, a black hole mass of 7.32$^{+0.89}_{-0.91}\times 10^{7}M_{\odot}$ from the $\sigma_{\mathrm{line}}$ of the rms spectrum is found for PKS0736.
\begin{table} 
	\centering
	\caption{Rest-frame resolution uncorrected line width and black hole mass measurements from mean and rms spectra.}
	\scalebox{0.8}{
		\begin{tabular}{lcccc}
			\hline\hline
			&Spectrum&Type&$\Delta V(\mathrm{km \, s^{-1}})$& $M_{\mathrm{BH}}$($\times 10^{7}M_{\odot}$)\\
			\hline\hline
			\multicolumn{5}{l}{}\\
			&Mean&FWHM&2160.14$^{+198.65}_{-210.48}$&6.78$^{+1.30}_{-1.26}$\\
			&&$\sigma_{\mathrm{line}}$&1230.92$^{+79.46}_{-74.03}$&8.78$^{+1.17}_{-1.02}$\\  
			&RMS&FWHM&1946.48$^{+284.58}_{-257.41}$&5.50$ ^{+1.72}_{-1.36}$\\ 
			&&$\sigma_{\mathrm{line}}$&1123.04$^{+66.58}_{-72.48}$&7.32$^{+0.89}_{-0.91}$\\  
			\hline
	\end{tabular}}
\begin{tablenotes}
\item \textbf{Notes.} Columns: Spectrum, line width type, line width, black hole mass.
\end{tablenotes}
\label{tab:Black-hole mass table}
\end{table}

\section{Discussion}
\label{Discussion}

\subsection{Effect of seasonal gaps on lag estimate}\label{sec:seasonal_gaps}
Due to the low declination, the monitoring observations of PKS0736 are affected by the seasonal gaps, as can be seen from the light curve plot. We have simulated light curves to investigate the impact of seasonal gaps on the lag measurement. We have constructed a mock continuum light curve using the DRW model of {\small JAVELIN} having the same properties (i.e., same amplitude and time-scale of variation) as the observed V-band continuum light curve. Then we simulated the emission line light curve from the mock continuum light curve with a shift of 80 days based on our calculated lag from the observed data, smoothing and scaling the mock light curve to mimic the observed light curve. We then down-sampled the light curves to have the same observational gaps.

We then performed all the time-series analysis methods to recover the time lag of 80 days from the simulated data. The results are shown in Fig. \ref{fig:Simulated Results} and tabulated in Table \ref{tab:Simulated lag table_A} for a single set of mock lightcurves. We noticed a strong peak at $\sim$80 days in all the methods, i.e., ICCF, {\small JAVELIN}, von Neumann, and Bartels, that well-recovered the input lag suggesting the observed cadence and seasonal gaps do not affect our lag estimation. Although a secondary peak of around 180 days has been seen from CCF and {\small JAVELIN}, it is well-separated from the primary peak and does not affect our lag measurements. Moreover, the peak of the ICCF ($r_{\mathrm{max}}\sim0.5$) matches well with that obtained from the observed data (see middle panel of Fig.\ref{fig:CCF plots}). We further simulated such 500 sets of continuum and line lightcurves and repeated the above process of measuring lag. In Fig. \ref{fig:simulated_tlaghist} we showed the distribution of lag ratio recovered from the ICCF to the input lag of 80 days, which provides a ratio of unity. Based on these simulations, we conclude a H$\beta$ lag of $\sim$80 days from the observed data of PKS0736. 

\subsection{Size-luminosity relation}\label{sec:size-luminoisty}
The PKS0736 is a radio-loud AGN having a combination of thermal emission from the disk and non-thermal emission from the jet elucidated by Fig. \ref{fig:NTD plot}. Consequently, the measurement of $L_{5100}$ is affected by non-thermal emission. However, we considered the quiescent state by removing the flaring region as visible in the $\gamma$-ray light curve (Fig. \ref{fig:lightcurves}). 

We plotted PKS0736 in the size-Luminosity diagram as shown in Fig. \ref{fig:size luminosity plot} along with various objects from the literature. The PKS0736 follows the best-fit size luminosity relation given by \citet{Bentz2013}.
Hence, its position is found to be consistent with the size$-$luminosity relation of other AGN. \citet{Du2019} has provided a new scaling relation taking into account the Fe II emission contamination in the spectrum of quasars and for especially high accreting AGN. This new scaling relation can explain the deviation seen in high-accreting sources from the size-luminosity relation. We estimate Fe II strength from the continuum subtracted mean spectrum. The strength of Fe II (R$_{FeII}$), which is the flux ratio of Fe II (4435-4685 \AA) to H$\beta$ is found to be 0.58$\pm$0.08. Therefore, PKS0736 is moderately strong Fe II emitter.  The time-lag estimated using scaling relation of \citet{Du2019} is around 63 light days, which is slightly lower than the 87 light days estimated using \citet{Bentz2013}.

The disk luminosity of PKS0736 was measured by several authors and found to be in the range of 10$^{44.6}$ to 10$^{45.7}$ ergs s$^{-1}$ \citep[see][]{Abdalla2020, Dai2007, McLure2001}. The scaling relation R$_{\mathrm{BLR}}$ = 10$^{17}$ $\sqrt{L_{\mathrm{disk}}/10^{45}}$cm \citep{2009MNRAS.397..985G}, provides R$_{\mathrm{BLR}}$ = 24-86 light-days, our measured BLR size of 66 light-days is within this range. Alternatively, the NTD in the quiescent state can be used as an estimator for removing non-thermal contribution from the observed L$_{5100}$. The median of NTD which is measured as 2.51$^{+1.52}_{-0.65}$ indicates  $\sim$33$\%$ of disk contribution in L$_{5100}$. If we use this revised L$_{5100}$ value of 1.44 $\times$ 10$^{44}$, the expected BLR size based on \citet{Bentz2013} is found to be 41 light-days, which is slightly smaller than our estimated rest frame BLR size of 66 light-days.

We have also performed the ICCF analysis between radio (see Fig. \ref{fig:lightcurves}) and V-band light curves. The correlation between radio and optical is found to be very weak with $r_{\mathrm{max}}$ $\sim$ 0.3. The measured lag is $\sim$ 255 light days which is much longer than the lag of H$\beta$. The weak correlation between radio and optical and a much longer radio lag than H$\beta$ lag suggest the radio emitting region is different than the H$\beta$ emitting region and radio is not the primary contributor in the V-band fluxes within the time-range studied.

\begin{figure}
\resizebox{9cm}{8cm}{\includegraphics{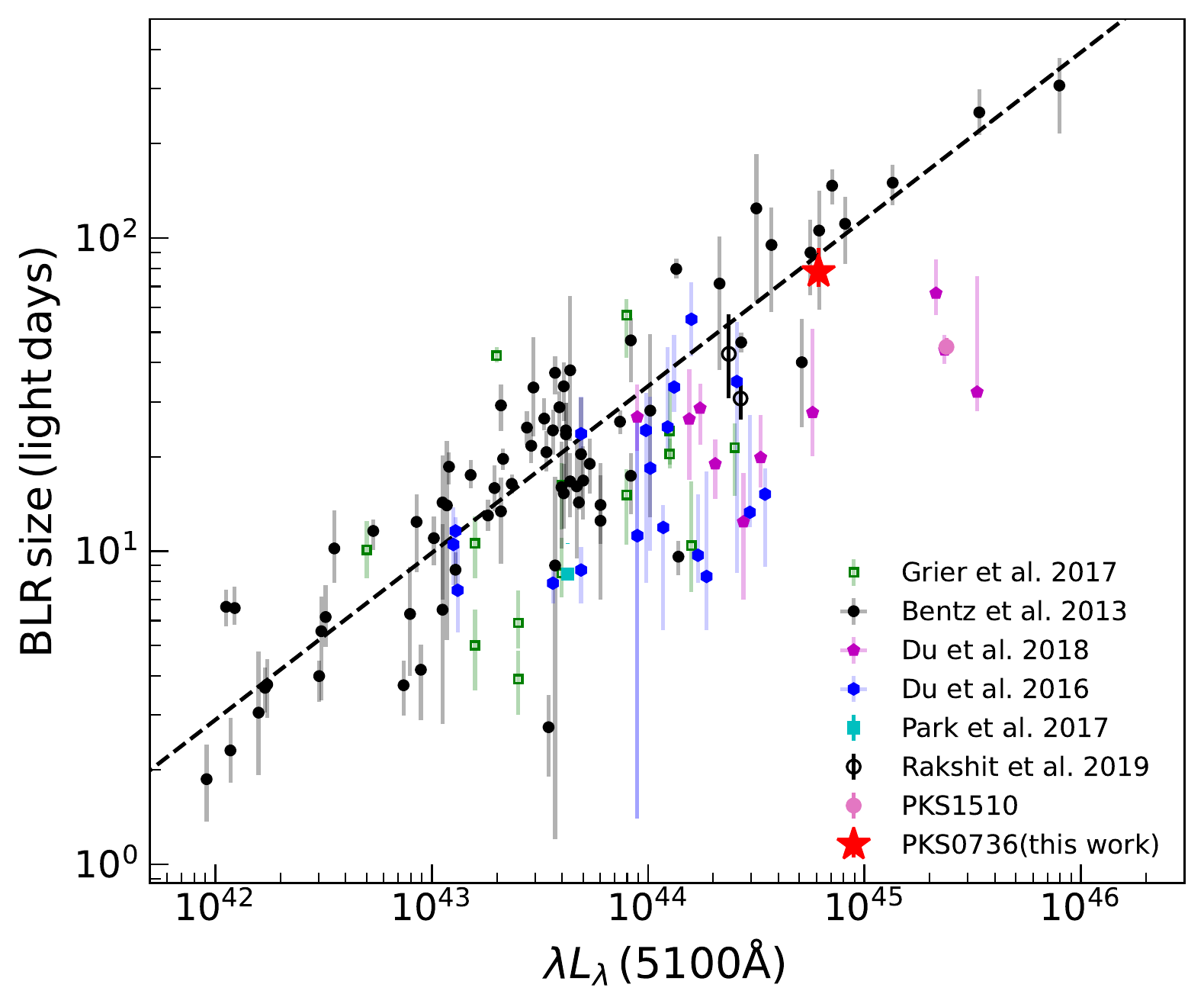}}
\caption{The plot is between the BLR size vs. $L_{5100}$ relation of AGNs. The source PKS0736 is following the relation very well. The best-fit relation of \citet{Bentz2013} is shown along with various RM results from the literature.}
\label{fig:size luminosity plot}
\end{figure}

\subsection{The effect of diffuse continuum}
Continuum reverberation mapping study of AGNs showed the UV-optical lags generally follow the $\tau\propto\lambda^{4/3}$ in agreement with the simple irradiated disk models; however, larger by a factor of 2-4 than predicated by standard accretion disk \citep[see, e.g.,][and references therein]{Fausnaugh_2016,Edelson_2019,2022arXiv220706432G}. Diffuse continuum (DC) emission from the BLR has been suggested as the possible origin of this larger-than-expected lag \citep{2001ApJ...553..695K,2018MNRAS.481..533L,2019NatAs...3..251C,2019MNRAS.489.5284K}. Recently, \citet{2022MNRAS.509.2637N} showed that time-dependent emission of the diffuse BLR gas could explain the observed large lag in the continuum RM. The DC could affect our NTD calculation and the BLR size-luminosity relation. Assuming the radiation pressure confined (RPC) cloud models and a covering factor of 0.2, the \citet{2022MNRAS.509.2637N} showed that the DC fraction could be 23\% of the total continuum at 5100A for a source with $L_{\mathrm{total, 5100}}=L_{\mathrm{disk, 5100}}+L_{\mathrm{DC, 5100}} =10^{44}$ erg s$^{-1}$.
The DC has two effects on the BLR size-luminosity relation. First, the true $L_{\mathrm{disk, 5100}}$ will be lower than the observed $L_{\mathrm{total, 5100}}$ and this will reduce the expected BLR size by 12\% ($\sim$4 days) for a  source with $L_{\mathrm{total, 5100}}=10^{44}$ erg s$^{-1}$ (see their equation 3). The NTD values are calculated based on the observed to predicted luminosity ratio. The $L_{5100, disk}$ should be lower if DC is taken into account; however, the predicted luminosity will also be affected as the correlation between L(H$\beta$) vs. $L_{\mathrm{total, 5100}}$ for the SDSS DR14 quasars is not corrected for DC. Second, since the lag is measured with respect to V-band, the reference point will be shifted toward a higher lag due to DC. Therefore, the true lag should be smaller by a few days, $\sim$4 days for $L_{\mathrm{total, 5100}}=10^{44}$ erg s$^{-1}$ \citep[see equation 6 of][]{2022MNRAS.509.2637N}. Both the size of the BLR and the luminosity of the continuum will be smaller by a similar extent for all the objects depending on the luminosity. Hence, the size-luminosity relation may not be affected much. A detailed quantitative analysis of this issue is beyond the scope of this paper.

\subsection{Black hole mass measurement}
Previous studies have estimated a range of black hole masses of PKS0736 from 10$^{8}$ to 10$^{8.73} M_{\odot}$ \citep{1992ApJ...398L..57S, Abdalla2020, Dai2007, Marchesini2004a, McLure2001}. These estimates are based on the scaling relationship (R$_{\mathrm{BLR}}$ $\propto$ $L_{5100}^{\alpha}$) \citep[e.g.,][]{Kaspi2000} measured from the single-epoch spectrum. Since PKS0736 is highly variable, the measurement of black hole mass from the single-epoch spectrum is highly uncertain. We note that single-epoch mass measurements are affected by choice of line width used to estimate the black hole. Generally, most of the single-epoch masses are calculated using FWHM; however, it has been found from the reverberation mapping study of AGNs with multiple emission lines that $\sigma_{\mathrm{line}}$ provides a better measure of the black hole mass \citep{Peterson_2004}. The rms spectrum isolates the non-varying components (e.g., narrow emission lines, host galaxy) and provides a robust estimation of black hole masses. Our reverberation-based black hole mass is found to be between 5.50-8.78 $\times 10^7 M_{\odot}$ depending on the line width estimator and type of spectrum. However, we consider the black hole masses of 7.32$^{+0.89}_{-0.91} \times 10^7 M_{\odot}$ calculated from the $\sigma_{\mathrm{line}}$ of the rms spectrum as the best black hole mass measurement.       

The bolometric luminosity of PKS0736 is calculated as 5.52 $\times$ 10$^{45}$ erg s$^{-1}$ using mean $L_{5100}$ and the relation $L_{\mathrm{BOL}}$ = 9 $\times$ $L_{5100}$ \citep{Kaspi2000}. The Eddington luminosity ($L_{\mathrm{EDD}}$) estimated as 9.22 $\times$ 10$^{45}$ erg s$^{-1}$ whereas the Eddington ratio  ($\lambda_{\mathrm{EDD}}$) is 0.60 calculated by $L_{\mathrm{EDD}}$ = 1.26 $\times$ 10$^{38} \,M _{\mathrm{BH}}$ and the black hole mass based on the $\sigma_{\mathrm{line}}$ of rms spectrum. This suggest that PKS0736 is accreting at a sub-Eddington rate. 

\section{Conclusions}
\label{Conclusions}
We analyzed the SO spectro-photometric data for PKS0736 obtained between November 2014 and May 2018, rendering more than 100 spectra and photometric data points. We found strong variability in the photometric light curve with $F_{\mathrm{var}}$=69.86$\pm$4.30\%, which is also reflected in the spectroscopic continuum with zero lag. Both the H$\beta$ and H$\gamma$ show high variability with $F_{\mathrm{var}}$ of 21.10$\pm$1.79\% and 29.73$\pm$ 2.25\%, respectively. The estimated NTD is found to be $2.51$ suggesting about 67\% contribution from the non-thermal emission towards the observed mean $L_{5100}$ of 6.13$\times10^{44}$ ergs s$^{-1}$. Using the cross-correlation method, we obtained a lag between optical V-band and H$\beta$ (H$\gamma$) of 66.4$^{+6.0}_{-4.2}$ (60.4$^{+9.2}_{-15.3}$) in the rest frame. Using virial relation and a scale factor of $f = 1.12$ ($f = 4.47$) for FWHM ($\sigma_{\mathrm{line}}$) a black hole mass of 7.32$^{+0.89}_{-0.91} \times 10^{7}M_{\odot}$ is obtained from the rms spectrum and $\sigma_{\mathrm{line}}$ of H$\beta$ line profile. The position of PKS0736 is consistent with the size$-$luminosity relation of AGNs.

\section*{Acknowledgements}
We thank the referee for comments and suggestions that helped to improve the quality of the manuscript. SR thanks Hagai Netzer for useful discussion about diffuse continuum. This publication makes use of data products from the Fermi Gamma-ray Space Telescope and accessed from the Fermi Science Support Center. Data from the Steward Observatory spectro-polarimetric monitoring project were used. This research has made use of data from the OVRO 40-m monitoring program \citep{2011ApJS..194...29R} which is supported in part by NASA grants NNX08AW31G, NNX11A043G, and NNX14AQ89G and NSF grants AST-0808050 and AST-1109911. SR acknowledges the partial support of SRG-SERB, DST, New Delhi through grant no. SRG/2021/001334. J.H.W. acknowledges funding from the Basic Science Research Program through the National Research Foundation of the Korean government (NRF- 2021R1A2C3008486).

\section*{Data Availability}
The data underlying this article were accessed from the Steward Observatory database with the following link: \url{http://james.as.arizona.edu/~psmith/Fermi/}. The derived data generated in this research are available in the article and in its online supplementary material.

\bibliographystyle{mnras}


\clearpage
\newpage
\appendix
\label{appendix}

\section{Analysis of total 107 spectra taken with 4.1$^{\prime\prime}$ slit}\label{sec:4arsec_mean_rms}
Although monitoring observations were carried out using multiple slits, the majority of them (84\% spectra) are taken with 4.1$^{\prime\prime}$ slit width. To investigate whether the different slit widths used during spectroscopic observations of PKS0736 have any effect on the line width measurement and subsequently the black hole mass estimation, we constructed the mean and rms spectra only with the 107 spectra taken with 4.1$^{\prime\prime}$ slit width.  
\begin{figure}
\resizebox{8cm}{8cm}{\includegraphics{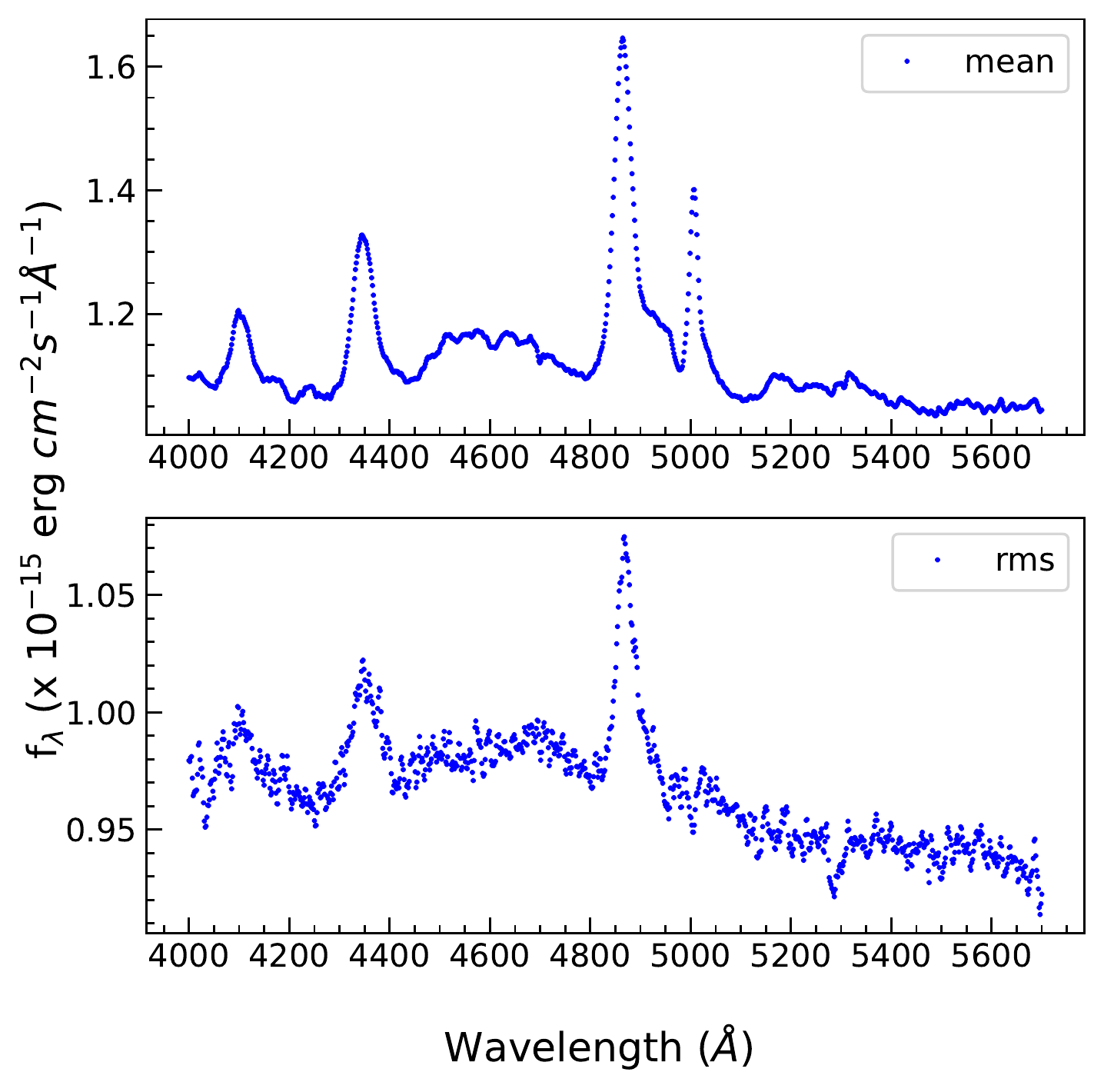}}
\caption{Mean and RMS spectra of PKS0736 from the 107 spectra obtained using 4.1$^{\prime\prime}$ slit after rescaling based on [O III]$\lambda$5007 without continuum subtracted from nightly spectra.}
\label{fig:Mean and RMS spectra_4arc}
\end{figure}
\begin{table} 
	\centering
	\caption{Rest-frame resolution uncorrected line width and black hole mass measurements from mean and rms spectra.}
		\begin{tabular}{lcccc}
			\hline\hline
			&Spectrum&Type&$\Delta V(\mathrm{km \, s^{-1}})$& $M_{\mathrm{BH}}$($\times 10^{7}M_{\odot}$)\\
			\hline\hline
			\multicolumn{5}{l}{}\\
			&Mean&FWHM&2147.08$^{+196.95}_{-222.35}$&6.60$^{+1.26}_{-1.29}$\\
			&&$\sigma_{\mathrm{line}}$&1223.05$^{+82.62}_{-73.72}$&8.55$^{+1.19}_{-1.00}$\\ 
			&RMS&FWHM&1934.60$^{+257.63}_{-258.21}$&5.36$^{+1.52}_{-1.33}$\\ 
			&&$\sigma_{\mathrm{line}}$&1064.81$^{+80.42}_{-79.15}$&6.48$^{+1.01}_{-0.92}$\\   
			\hline
	\end{tabular}
\begin{tablenotes}
\item \textbf{Notes.} Columns: Spectrum, line width type, line width, black hole mass.
\end{tablenotes}
\label{tab:Black-hole mass table_B}
\end{table}
Fig. \ref{fig:Mean and RMS spectra_4arc} is remarkably similar to the mean and rms spectra obtained from entire spectroscopic data (see Fig. \ref{fig:Mean and RMS spectra}) provided from SO irrespective of the slit width used during observation. Moreover, we estimated FWHM and $\sigma$ from these mean and rms spectra. The results are given in Table \ref{tab:Black-hole mass table_B}, which shows consistent results with that obtained with entire spectroscopic data (see Table \ref{tab:Black-hole mass table}). It suggests that the use of entire spectral data do not affect black hole mass measurement. 

\section{Time series analysis of simulated light curves}
We have performed time-series analysis on the simulated data as mentioned in section \ref{sec:seasonal_gaps}. The results of the time-series analysis are shown in Fig. \ref{fig:Simulated Results} for ICCF, {\small JAVELIN}, Von-Neumann, and Bartels. Table \ref{tab:Simulated lag table_A} summarizes the results. 

\begin{figure}
\resizebox{6cm}{4cm}{\includegraphics{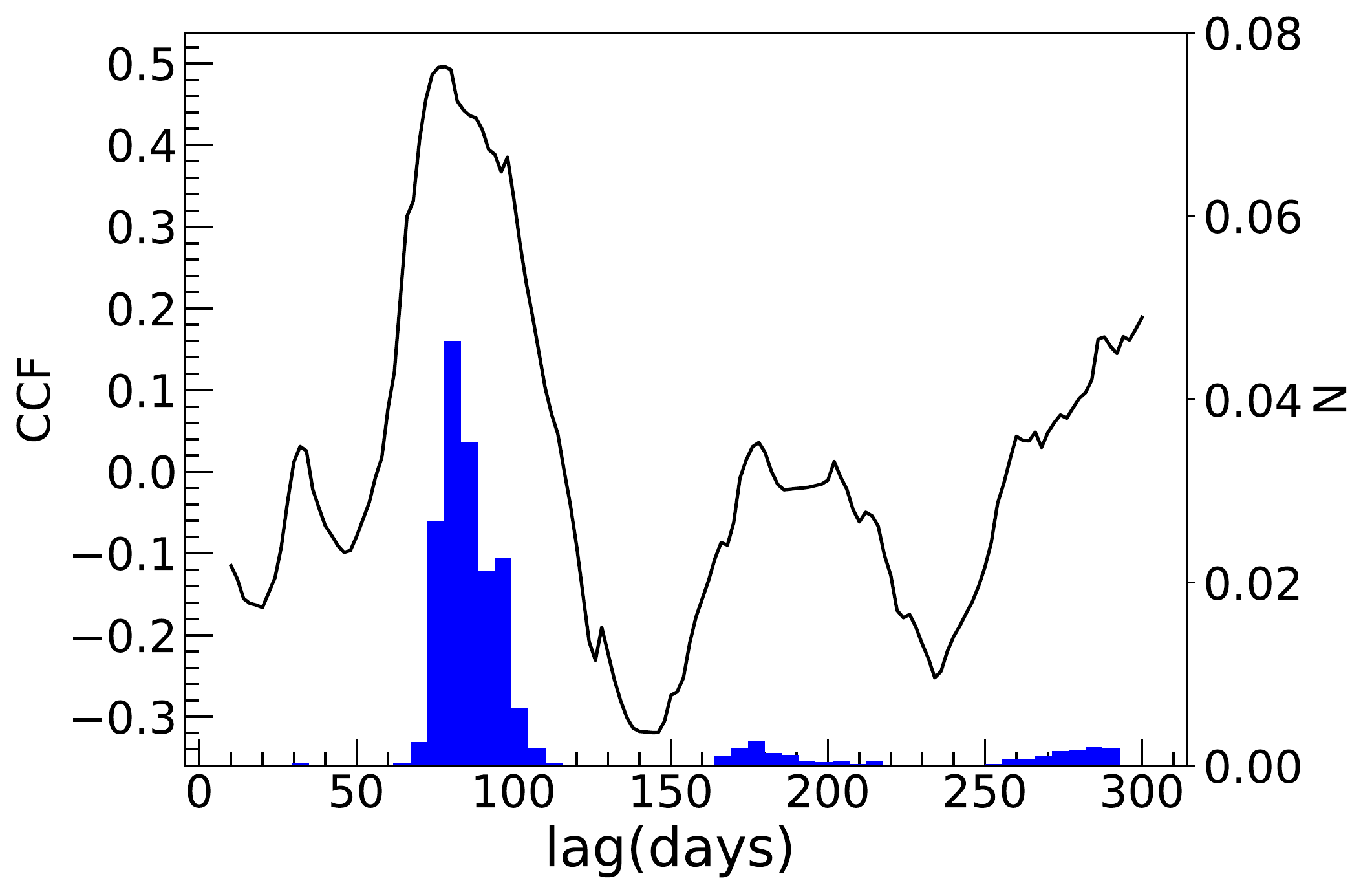}}
\resizebox{4cm}{4cm}{\includegraphics{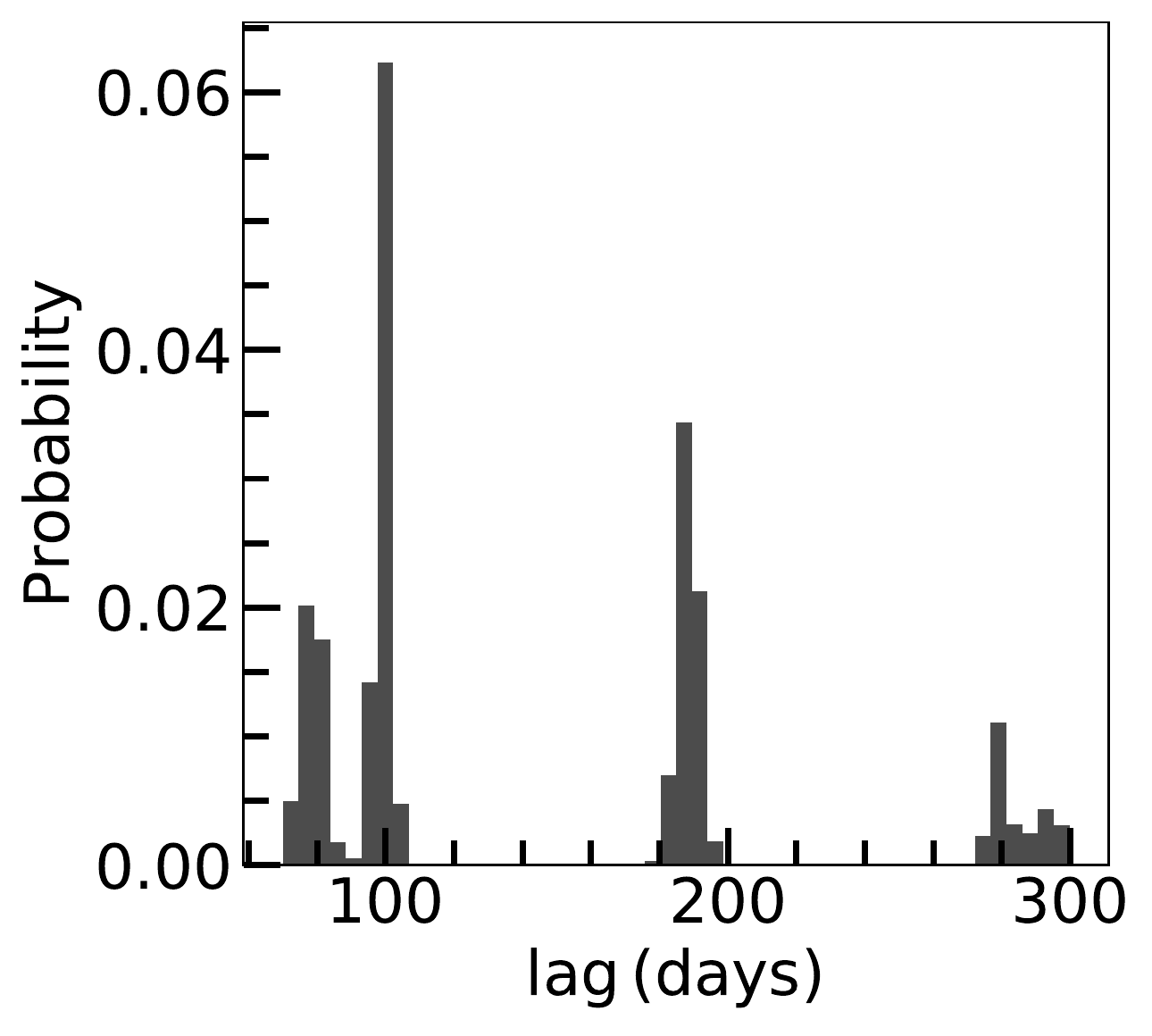}}
\newline
\resizebox{8cm}{4cm}{\includegraphics{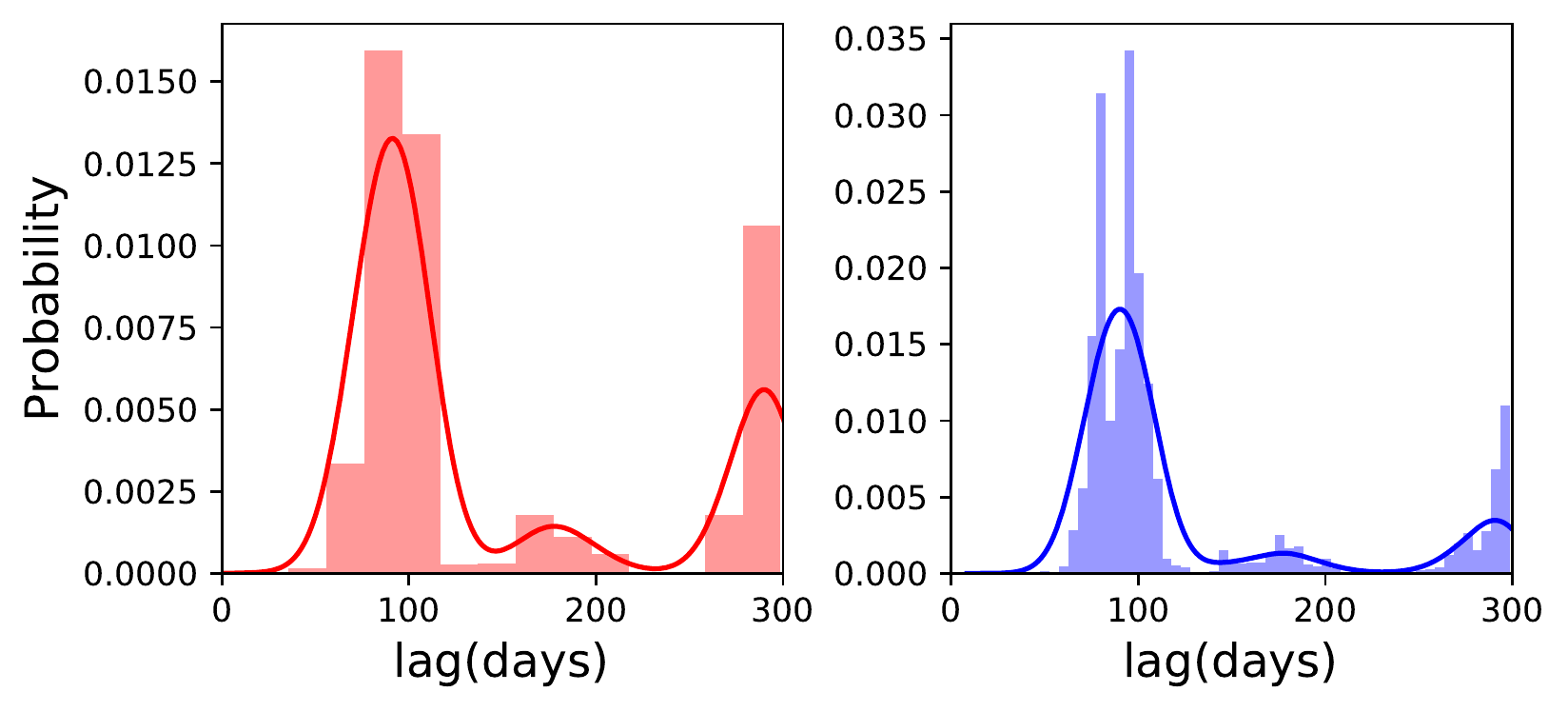}}
\caption{Lag estimation from the simulated light curves. Top: Cross-correlation function and lag distribution for simulated data. The ICCF (line) and centroid probability distribution from ICCF (filled histogram) are shown. Middle: Lag probability distribution obtained from {\small JAVELIN} for simulated data. Bottom: Lag probability distribution obtained from Von-Neumann (left) and Bartels (right) for simulated data.}
\label{fig:Simulated Results}
\end{figure}

\begin{table}
	\centering
    \begin{tabular}{lc} 
		\hline
		Method & Lag (days) \\
		 \hline
		ICCF & 84.0$^{+7.1}_{-11.0}$ \\
		JAVELIN & 97.7$^{+2.8}_{-17.7}$ \\
		von Neumann & 90.4$^{+13.4}_{-13.1}$\\
		Bartels &  90.5$^{+12.6}_{-11.5}$ \\
		\hline
	\end{tabular}
	\caption{Columns as follows: (1) the method used to calculate the lag (days), (2) lag recovered from the simulated light curves with an input lag of 80 days.}
	\label{tab:Simulated lag table_A}
\end{table}
\begin{figure}
\flushleft
\resizebox{8.5cm}{6cm}{\includegraphics{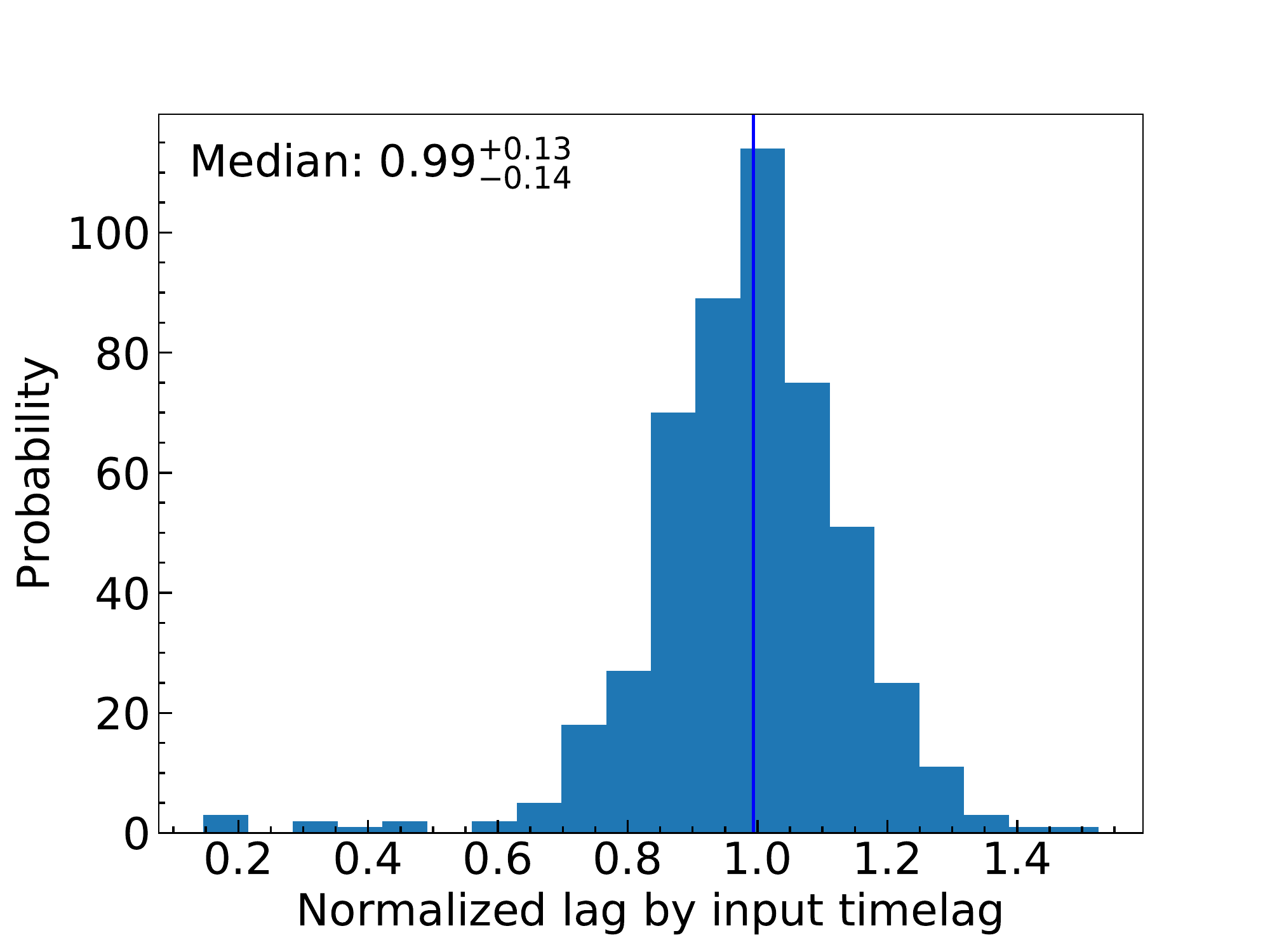}}
\caption{We simulated the light curves for both photometric V-band flux and H$\beta$ flux over 500 times and plotted the distribution of the ratio of the output lag to input lag (80 days). The median of the histogram which is closer to value 1 suggests the estimated lag is reliable.}
\label{fig:simulated_tlaghist}
\end{figure}

\bsp	
\label{lastpage}
\end{document}